\documentclass[10pt,Journal]{IEEEtran}\IEEEoverridecommandlockouts
\usepackage{algorithmic}
\usepackage[linesnumbered,ruled]{algorithm2e}
\usepackage{amsfonts}
\usepackage{amsmath}
\usepackage{amssymb}
\usepackage{amsthm}
\usepackage{bm}
\usepackage{bbm}
\usepackage{booktabs}
\usepackage{caption}
\usepackage{CJK}
\usepackage{color}
\usepackage{cuted}
\usepackage{geometry}
\usepackage{graphicx}
\usepackage{indentfirst}
\usepackage[marginal]{footmisc}
\usepackage{multirow}

\allowdisplaybreaks[4]

{\bgroup
 \addtolength\abovedisplayshortskip{#1}
 \addtolength\abovedisplayskip{#1}
 \addtolength\belowdisplayshortskip{#1}
 \addtolength\belowdisplayskip{#1}}
{\egroup\ignorespacesafterend}

\newcounter{MYtempeqncnt}

\theoremstyle{plain}
\newtheorem{thm}{Theorem}

\begin{document}

\title{Channel Estimation By Transmitting Pilots From Reconfigurable Intelligent Surface}

\author{Yanze Zhu, Yang Liu, \thanks{\hspace{0.35cm}Y. Zhu and Y. Liu are with the School of Information and Communication Engineering, Dalian University of Technology, Dalian, China, email: 5476z4969y5283z@mail.dlut.edu.cn, yangliu\_613@dlut.edu.cn.} Qingqing Wu, \thanks{\hspace{0.35cm}Q. Wu is with the Department of Electronic Engineering, Shanghai Jiao Tong University, 200240, China, email: qingqingwu@sjtu.edu.cn.} Changsheng You, \thanks{\hspace{0.35cm}C. You is with the Department of Electronic and Electrical Engineering, Southern University of Science and Technology, Shenzhen, China, email: youcs@sustech.edu.cn.} and Qingjiang Shi \thanks{\hspace{0.35cm}Q. Shi is with the School of Software Engineering, Tongji University, Shanghai, China, and also with the Shenzhen Research Institute of Big Data, Shenzhen, China, email: shiqj@tongji.edu.cn.}
\vspace{-0.2cm}}

\maketitle

\begin{abstract}
Reconfigurable intelligent surface (RIS) is a promising technology for future wireless communication systems. Channel estimation (CE) of RIS device is a critical but also challenging issue for its development. The mainstream of existing CE methods is confined to the so-called cascaded channel (CscdChn) estimation scheme, which treats the multiplicative two-hop RIS channels as an effective one and measures it as a whole. This CscdChn training method suffers from severe double-fading attenuation loss, which significantly degrades the CE accuracy. In this paper, we propose a novel RIS-transmitting (RIS-TX) based CE scheme, which has lower pilot overhead than CscdChn scheme and effectively overcomes the double-fading curse via incorporating only one single transmit radio frequency (RF)-chain into RIS. We develop highly efficient gradient descent (GD) and penalty duality decomposition (PDD)-based solutions to resolve the pilot design task for the RIS-TX CE scheme, which is a difficult quartic optimization problem. Our designed pilot signal outperforms the discrete Fourier transform (DFT) sequence, which is reported to be optimal for CscdChn scheme. Besides, both theoretical analysis and numerical results demonstrate that our proposed RIS-TX scheme exhibits distinct performance characteristics as opposed to its CscdChn counterpart and yields superior accuracy when RIS device is not extremely large.
\end{abstract}


\section{Introduction}

\subsection*{A. Background}

The emerging reconfigurable intelligent surface (RIS) [1], which is also known as intelligent reflecting surface (IRS) [2], has attracted upsurging interests from both academia and industry, and is envisioned as a promising technology for the next-generation wireless communication systems [3]. The RIS can boost the network communication performance from various aspects while consuming rather low energy. Potentials of RIS have been extensively explored recently and many exciting applications can be found in [1]-[3] and the references therein.

Besides its versatility, one most critical aspect of RIS technology is the channel state information (CSI) acquisition. CSI is essential to configure RIS' phase-shifting, which, however, is practically difficult to obtain for RIS due to its incapability of transmitting, receiving or processing signals. Therefore, a multitude of recent works are dedicated to exploring channel estimation (CE) techniques suitable for RIS, e.g., the references in [4]. Among them, most available methods belong to the so-called cascaded channel (CscdChn) estimation scheme, e.g., [5]-[13]. That is, the effective channel cascading the two-hop links by way of RIS is measured as a whole during the training procedure. For instance, the work [5] proposes to turn on one single RIS element at a time and estimate each CscdChn one after another. In [6], the authors design the pilot sequence according to the minimum variance unbiased estimation rule. The authors of [7] decompose the CscdChn into rank-one subchannels and design pilot signals via adjusting RIS phase-shifting to minimize the mean square error (MSE) of the estimate of subchannels. The paper [8] proposes both linear minimum mean square error (LMMSE) based and deep learning based pilot design methods to improve the CE precision of the CscdChn and shows that discrete Fourier transform (DFT) pilots can achieve nearly optimal MSE. The authors of [9] wisely exploit the fact that the link between the base station (BS) and RIS are the same for all users' CscdChns to reduce the overall training overhead of multi-user system. The work [10] studies the RIS training method for orthogonal frequency division multiplexing (OFDM) system via optimizing the training phase-shifts and demonstrates that the DFT sequence can achieve the optimal performance. The authors of [11] propose a hierarchical training scheme via grouping the RIS elements and performing CE progressively using discrete phase shifters. Besides, a line of research, e.g., [12] and [13], explores to leverage the hidden sparsity in channel for high frequency/large antenna array systems to decrease training overhead.

Although the CscdChn training method predominates the existing literature, there still exist a number of works exploring \emph{active} methods at the RIS side for channel training [14]-[18]. For example, the work [14] proposes a hybrid RIS structure via embedding sensors in element array to measure channels. The authors of [15] install one single receive (RX) RF-chain in RIS and utilize it to develop a compressive sensing (CS) based channel recovery algorithm via exploiting channel's sparsity. In [16], active sensing elements are installed on the uniform planar element array of the RIS and arranged in “L” shape to sense the channels. The latest works [17] and [18] consider similar hybrid RIS architecture as proposed in [14] and utilize tensor decomposition and atomic norm methods to recover the sparse channels, respectively. Besides, some works investigate extraction of one-hop CSI of BS-RIS or RIS-user through the reflected pilot signals. For instance, the authors of [19] and [20] propose factor decomposition and matrix completion methods to achieve this goal. The work [21] exploits matrix calibration theory and channel sparsity to recover separate RIS channels.

\subsection*{B. Motivation}

Along with the deepening research on RIS technology, some critical insights have been obtained very recently. Especially, as pointed out by [22], one predominant defect of RIS is the severe fading loss of the concatenated channels. This feature is named as \emph{double-fading} effect in [22] and derives from the multiplicative nature of the CscdChn's attenuation. As analyzed in [22], the attenuation loss experienced by the two-hop channels going through RIS is generally several orders of magnitude larger than that of the direct channel. This effect has also been substantiated by numerical experiments and real-field tests reported in the latest works [23] and [24]. At the same time, according to Cram\'er-Rao lower bound (CRLB) analysis presented in [1] and [25], the MSE of CE is inversely proportional to the signal-to-noise-ratio (SNR). Therefore, the double-fading loss indeed severely weakens the pilot signals' receiving SNR and hence bottlenecks the CE performances of the classical CscdChn training schemes, e.g., [5]-[13], and the one-hop CSI extraction methods relying on the reflected pilot signals, e.g., [19]-[21].

Based on the above inspections, to improve the CE precision, one natural thought is to breakthrough the double-fading curse. In fact, a number of works have made efforts toward this end. For instance, the authors of [23] and [24] have proposed a novel \emph{active}-RIS architecture by introducing amplifiers into RIS elements to magnify the impinging signals. At the same time, the works [14]-[18] introduce a small number of RX RF-chains at the RIS side to perform signal receiving locally.

Enlightened by all the above works, this paper proposes a novel RIS channel training scheme. Specifically, we equip the RIS with one single TX RF-chain [15], [26], as shown in Fig. 1. During the channel training period, the TX RF-chain is connected to all the phase shifters [26], which enables the RIS to broadcast pilot signals to the BS and all users. This novel scheme cherishes the following advantages:

\begin{itemize}

\item This scheme can overcome the double-fading curse. In fact, only “one-hop” channels are estimated.

\item Only one TX RF-chain is required, which can largely maintain the hardware efficiency of RIS device.

\item The BS and all users conduct CE simultaneously. This makes the RIS channel training overhead independent of the number of users.

\end{itemize}

\begin{figure} [!t]
\centering
\includegraphics[scale=0.30]{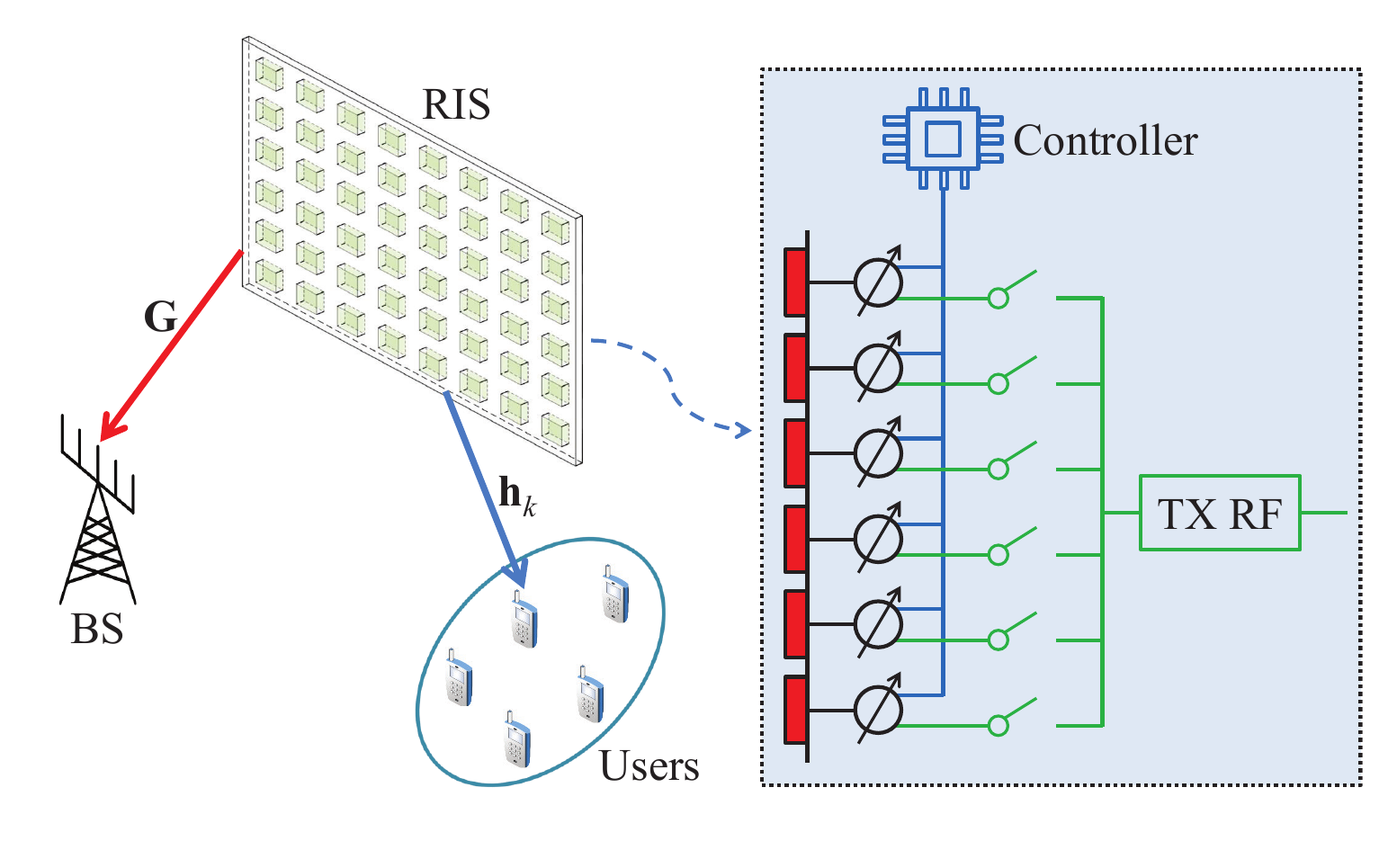}
\caption*{Fig. 1. CE based on the proposed transmit-able RIS.}
\end{figure}

\subsection*{C. Contributions}

Motivated by the above observations, this paper proposes a novel RIS-transmitting (RIS-TX) channel training scheme as specified above. We conduct a comprehensive research on this new scheme. Specifically, the contributions of this paper are specified as follows:

\begin{itemize}

\item Firstly, we put forward a brand new RIS-TX CE scheme. Via incorporating one TX RF-chain into RIS device, we propose that RIS broadcasts “pilot signals”, which are indeed implemented by RIS elements' predefined phase shifting, to the BS and all mobile users. This novel CE scheme's pilot overhead is lower than the conventional CscdChn methods, e.g., [4]-[11], [19], [20], and can effectively overcome double-fading effect. To the best of authors' knowledge, similar RIS-TX scheme has never been considered in the existing literature, e.g., [4]-[26].

\item Next, we study the associated pilot sequence design problem. Specifically, we aim at minimizing the estimation MSE of the effective RIS channels of all users via designing the pilot sequences. This task turns out to be a nonconvex quartic optimization problem and is much more challenging than those in the existing relevant literature, e.g., [6]-[11]. To tackle this difficult problem, we successfully develop a gradient descent (GD)-based solution, which exhibits fast convergence.

\item Besides, we also develop an alternative pilot design solution by exploiting the cutting-the-edge penalty duality decomposition (PDD) framework [27]. This PDD-based solution has all its block coordinates updated either in closed form or by solving a structured linear equation and consequently runs highly efficiently.

\item Furthermore, we also theoretically analyze the performance of our RIS-TX training scheme. CRLB analysis has uncovered one important fact that the number of RIS elements exerts opposite impact on the RIS-TX scheme and the prevailing CscdChn counterpart. The former yields superior estimation accuracy when RIS has moderate size (e.g. $\leq$ 10000 elements) while the latter is more beneficial for large-scale RIS. Asymptotic analysis reflects that the two competing CE schemes exhibit distinct scaling laws in power and RIS size. These insights provide valuable guidelines for selecting between the RIS-TX and CscdChn scheme in various application scenarios, which have never been discussed in the existing literature, e.g., [4]-[24].

\item Last but not least, extensive numerical results demonstrate the effectiveness of our proposed algorithms and the benefit of the novel RIS-TX training scheme. Very interestingly, the pilot sequences optimized by our GD/PDD-based algorithms outperform the DFT pilot sequence, which has been reported to be optimal for the CscdChn scheme by different literature [6]-[8], [10], [11].

\end{itemize}

\section{System Model and Problem Formulation}

In this section, we will elaborate the system setting, the channel training procedure and introduce the problem formulation of pilot design towards minimizing the CSI estimation errors.

\subsection*{A. System Setting}

As illustrated in Fig. 1, we consider a narrowband time division duplex (TDD) cellular system which consists of a BS equipped with $M_{\mathrm{B}}$ antennas, $K$ single-antenna users and a RIS containing $M_{\mathrm{R}}$ reflecting elements. The sets of users and RIS elements are denoted by $\mathcal{K} \triangleq \{ 1, \cdots, K \}$ and $\mathcal{M} \triangleq \{ 1, \cdots, M_{\mathrm{R}} \}$, respectively. Moreover, we denote $\mathbf{G} \in \mathbb{C}^{M_{\mathrm{B}} \times M_{\mathrm{R}}}$ and $\mathbf{h}_{k} \in \mathbb{C}^{M_{\mathrm{R}}}, \; k \in \mathcal{K}$, as the channel between the BS and the RIS and the one between the RIS and the $k$th user, respectively. Note that our setting does not consider the direct channels connecting the BS and users for simplicity, which can be easily acquired by following the standard CE procedure for conventional cellular system without RIS [28].

As previously discussed, one RF-chain is built in the RIS device to empower it with transmission capability, as shown in Fig. 1. During the channel training period, every RIS element is switched to connecting the single RF-chain and adjusts the signal's phase before emitting it over the air. Specifically, denote the RF-chain signal as a complex scalar $\sqrt{\alpha}e^{j\theta_{0}}$ with $\sqrt{\alpha}$ and $\theta_{0}$ representing its amplitude and phase, respectively. Therefore, the emitted signal from the RIS can be expressed as $\bm{\psi} \triangleq [\sqrt{\alpha}e^{j(\theta_{0} + \theta_{1})}, \cdots, \sqrt{\alpha}e^{j(\theta_{0} + \theta_{M_{\mathrm{R}}})}]^\mathrm{T}$ with $\theta_{i}, \; i \in \mathcal{M}$, being the phase shift yielded by the $i$th element. Since the pilot signal is predefined, we can just assume $\theta_{0} = 0$ without loss of generality.

\vspace{-4mm}
\subsection*{B. Channel Estimation}

Based on the above, we elaborate the proposed RIS-TX CSI acquisition procedure. The training period is composed of $T$ consecutive time slots and is assumed to be shorter than the channel's coherence time. Within each time slot, the RIS broadcasts one pilot symbol, which is phase-modulated by the RIS elements, to the BS and all users. Specifically, denoting $\mathcal{T} \triangleq \{ 1, \cdots, T \}$, we can express the transmitted signal in the $t$th time slot as [7]-[11]
\begin{equation}
\mathbf{x}^{(t)} = \bm{\psi}^{(t)} = \sqrt{\alpha_{t}}\bm{\phi}^{(t)}, \; t \in \mathcal{T},
\end{equation}
where $\bm{\phi}^{(t)} \triangleq [e^{j\theta_{1}^{(t)}}, \cdots, e^{j\theta_{M_{\mathrm{R}}}^{(t)}}]^\mathrm{T}$.

After collecting all $T$ pilot symbols from the RIS simultaneously, the BS and all users conduct signal processing and obtain the estimate of their own channel connecting the RIS. This procedure will be detailed in the sequel.

In the $t$th time interval, the BS receives the signal from the RIS which is given as
\begin{equation}
\mathbf{y}_{\mathrm{B}}^{(t)} = \mathbf{G}\bm{\psi}^{(t)} + \mathbf{n}_{\mathrm{B}}^{(t)}, \; t \in \mathcal{T},
\end{equation}
where $\mathbf{n}_{\mathrm{B}}^{(t)}, \; t \in \mathcal{T}$, represents the thermal noise at the BS and $\mathbf{n}_{\mathrm{B}}^{(t)} \sim \mathcal{CN}(\bm{0}, \bm{\Sigma}_{\mathrm{B}})$ \textcolor{blue}{with the positive definite matrix $\bm{\Sigma}_{\mathrm{B}}$ denoting the covariance matrix of the noise which is assumed to be known [8], [9], [29].$^{1}$\footnote{$^{1}$\textcolor{blue}{Note the noise statistics generally vary slowly in a large timescale [30], which hence can be obtained at a low expense of time overhead. It is also reasonable to assume that the noise covariance matrix is invariant during the coherence time, which indicates that different $\mathbf{n}_{\mathrm{B}}^{(t)}$'s have one same covariance matrix, i.e., $\bm{\Sigma}_{\mathrm{B}}$.}}} Besides, we assume that $\mathbf{n}_{\mathrm{B}}^{(t)}$'s with different $t$ are independent for simplicity [6]-[11]. To fully determine the channel $\mathbf{G}$, the length of the pilot sequence $T$ should satisfy $T \geq M_{\mathrm{R}}$. To reduce training's overhead, we just assume that $T = M_{\mathrm{R}}$. By collecting all $T$ observations and stacking them in a column-by-column manner, the training data obtained by the BS can be represented as
\begin{equation}
\mathbf{Y}_{\mathrm{B}} = \mathbf{G}\bm{\Psi} + \mathbf{N}_{\mathrm{B}},
\end{equation}
where $\mathbf{Y}_{\mathrm{B}} \triangleq [\mathbf{y}_{\mathrm{B}}^{(1)}, \cdots, \mathbf{y}_{\mathrm{B}}^{(T)}]$, $\bm{\Psi} \triangleq [\bm{\psi}^{(1)}, \cdots, \bm{\psi}^{(T)}]$, $\mathbf{N}_{\mathrm{B}} \triangleq [\mathbf{n}_{\mathrm{B}}^{(1)}, \cdots, \mathbf{n}_{\mathrm{B}}^{(T)}]$. Via vectorizing $\mathbf{Y}_{\mathrm{B}}$ and utilizing the formula $\mathsf{vec}(\mathbf{ABC}) = (\mathbf{C}^{\mathrm{T}} \otimes \mathbf{A})\mathsf{vec}(\mathbf{B})$, we obtain
\begin{equation}
\mathbf{y}_{\mathrm{B}} = \mathsf{vec}(\mathbf{Y}_{\mathrm{B}}) = (\bm{\Psi}^{\mathrm{T}} \otimes \mathbf{I}_{M_{\mathrm{B}}})\mathbf{g} + \mathbf{n}_{\mathrm{B}} = \mathsf{A}_{\mathrm{G}}(\bm{\Psi})\mathbf{g} + \mathbf{n}_{\mathrm{B}},
\end{equation}
where $\mathbf{g} \triangleq \mathsf{vec}(\mathbf{G})$, $\mathbf{n}_{\mathrm{B}} \triangleq \mathsf{vec}(\mathbf{N}_{\mathrm{B}})$, $\mathsf{A}_{\mathrm{G}}(\bm{\Psi}) \triangleq \bm{\Psi}^{\mathrm{T}} \otimes \mathbf{I}_{M_{\mathrm{B}}}$ and $\mathbf{I}_{M_{\mathrm{B}}}$ is $M_{\mathrm{B}} \times M_{\mathrm{B}}$ identity matrix.

Similarly, the compact form of the signals observed by the $k$th user during the whole training period can be expressed as
\begin{equation}
\mathbf{y}_{\mathrm{U},k} = \bm{\Psi}^{\mathrm{T}}\mathbf{h}_{k} + \mathbf{n}_{\mathrm{U},k}, \; k \in \mathcal{K},
\end{equation}
where $\mathbf{n}_{\mathrm{U},k} \triangleq [n_{\mathrm{U},k}^{(1)}, \cdots, n_{\mathrm{U},k}^{(T)}]^{\mathrm{T}}$ and $n_{\mathrm{U},k}^{(t)}, \; k \in \mathcal{K}, \; t \in \mathcal{T}$, indicates the thermal noise at the $k$th user in the $t$th interval with $n_{\mathrm{U},k}^{(t)} \sim \mathcal{CN}(0, \sigma_{\mathrm{U},k}^{2})$. For simplicity, it is assumed that each $n_{\mathrm{U},k}^{(t)}$ is independent to others for different users and time slots.

In this paper, we assume that the channels $\mathbf{g}$ and $\{ \mathbf{h}_{k} \}_{k=1}^{K}$ have zero mean and known covariance matrices. Specifically, we denote $\mathbf{R}_{\mathrm{G}}$ and $\mathbf{R}_{\mathrm{h},k}, \; k \in \mathcal{K}$, as covariance matrices of $\mathbf{g}$ and $\mathbf{h}_{k}, \; k \in \mathcal{K}$, respectively. In reality, channel statistics generally vary slowly and can be easily obtained [29]. For instance, the covariance matrix can be empirically estimated at low overhead [31], [32]. Besides, the zero mean assumption can also be readily satisfied via subtracting their nonzero mean values. Besides, it is reasonable to assume that the channels associated with the BS and users are mutually uncorrelated. With the linear receivers $\mathbf{W}_{\mathrm{G}}$ at the BS and $\mathbf{W}_{\mathrm{h},k}, \; k \in \mathcal{K}$, at the $k$th user, the BS/every user obtains the linear estimate of their own channels given as follows
\begin{align}
&\hat{\mathbf{g}} = \mathbf{W}_{\mathrm{G}}^{\mathrm{H}}\mathbf{y}_{\mathrm{B}} = \mathbf{W}_{\mathrm{G}}^{\mathrm{H}}\mathsf{A}_{\mathrm{G}}(\bm{\Psi})\mathbf{g} + \mathbf{W}_{\mathrm{G}}^{\mathrm{H}}\mathbf{n}_{\mathrm{B}},\\
&\hat{\mathbf{h}}_{k} = \mathbf{W}_{\mathrm{h},k}^{\mathrm{H}}\mathbf{y}_{\mathrm{U},k} = \mathbf{W}_{\mathrm{h},k}^{\mathrm{H}}\bm{\Psi}^{\mathrm{T}}\mathbf{h}_{k} + \mathbf{W}_{\mathrm{h},k}^{\mathrm{H}}\mathbf{n}_{\mathrm{U},k}, \; k \in \mathcal{K},
\end{align}
whose covariance matrices are given as follows after some manipulations
\begin{align}
&\mathbf{C}_{\hat{\mathrm{G}}} \!=\! \mathbf{W}_{\mathrm{G}}^{\mathrm{H}}\mathsf{A}_{\mathrm{G}}(\bm{\Psi})\mathbf{R}_{\mathrm{G}}\mathsf{A}_{\mathrm{G}}^{\mathrm{H}}(\bm{\Psi})\mathbf{W}_{\mathrm{G}} \!+\! \mathbf{W}_{\mathrm{G}}^{\mathrm{H}}(\bm{\Sigma}_{\mathrm{B}} \!\!\otimes\!\! \mathbf{I}_{M_{\mathrm{R}}})\mathbf{W}_{\mathrm{G}},\\
&\mathbf{C}_{\hat{\mathrm{h}},k} \!=\! \mathbf{W}_{\mathrm{h},k}^{\mathrm{H}}\bm{\Psi}^{\mathrm{T}}\mathbf{R}_{\mathrm{h},k}\bm{\Psi}^{\mathrm{*}}\mathbf{W}_{\mathrm{h},k} \!+\! \mathbf{W}_{\mathrm{h},k}^{\mathrm{H}}\bm{\Sigma}_{\mathrm{U},k}\mathbf{W}_{\mathrm{h},k}, k \!\in\! \mathcal{K},
\end{align}
where $\bm{\Sigma}_{\mathrm{U},k} \triangleq \mathsf{Diag}(\sigma_{\mathrm{U},k}^{2}, \cdots, \sigma_{\mathrm{U},k}^{2}), \; k \in \mathcal{K}$.

\vspace{-4mm}
\subsection*{C. Pilot Design Problem}

After performing CE, mobile users feedback their local estimates to the BS. The effective CscdChn by way of the RIS for each user is defined by $\mathbf{H}_{\mathrm{c},k} \triangleq \mathbf{G}\mathsf{Diag}(\mathbf{h}_{k}), \; k \in \mathcal{K}$, and its estimate is given as $\hat{\mathbf{H}}_{\mathrm{c},k} \triangleq \hat{\mathbf{G}}\mathsf{Diag}(\hat{\mathbf{h}}_{k}), \; k \in \mathcal{K}$. Based on $\{ \hat{\mathbf{H}}_{\mathrm{c},k} \}_{k=1}^{K}$, the BS will determine the scheduling and beamforming for all users. Therefore, it is critical to minimize the estimation error between the estimate $\{ \hat{\mathbf{H}}_{\mathrm{c},k} \}_{k=1}^{K}$ and their true values. To evaluate the estimation performance, the MSE matrix of $\hat{\mathbf{H}}_{\mathrm{c},k}, \; k \in \mathcal{K}$, is defined as $\mathbf{C}_{\mathrm{H}_{\mathrm{c}},k} \triangleq \mathbb{E} \{ \mathsf{vec}(\mathbf{H}_{\mathrm{c},k} - \hat{\mathbf{H}}_{\mathrm{c},k})\mathsf{vec}^{\mathrm{H}}(\mathbf{H}_{\mathrm{c},k} - \hat{\mathbf{H}}_{\mathrm{c},k}) \}, \; k \in \mathcal{K}$, and its value is determined by the following theorem, which is derived in Appendix A.

\begin{thm}

The value of $\mathbf{C}_{\mathrm{H}_{\mathrm{c}},k} \triangleq \mathbb{E} \{ \mathsf{vec}(\mathbf{H}_{\mathrm{c},k} - \hat{\mathbf{H}}_{\mathrm{c},k})\mathsf{vec}^{\mathrm{H}}(\mathbf{H}_{\mathrm{c},k} - \hat{\mathbf{H}}_{\mathrm{c},k}) \}, \; k \in \mathcal{K}$, is given by
\begin{align}
&\mathbf{C}_{\mathrm{H}_{\mathrm{c}},k} = \mathbf{R}_{\mathrm{G}} \odot (\mathbf{R}_{\mathrm{h},k} \otimes \mathbbm{1}_{M_{\mathrm{B}}}) + \mathbf{C}_{\hat{\mathrm{G}}} \odot (\mathbf{C}_{\hat{\mathrm{h}},k} \otimes \mathbbm{1}_{M_{\mathrm{B}}}) \notag\\
&- (\mathbf{R}_{\mathrm{G}}\mathsf{A}_{\mathrm{G}}^{\mathrm{H}}(\bm{\Psi})\mathbf{W}_{\mathrm{G}}) \odot ((\mathbf{R}_{\mathrm{h},k}\bm{\Psi}^{\mathrm{*}}\mathbf{W}_{\mathrm{h},k}) \otimes \mathbbm{1}_{M_{\mathrm{B}}}) \notag\\
&- ((\mathbf{R}_{\mathrm{G}}\mathsf{A}_{\mathrm{G}}^{\mathrm{H}}(\bm{\Psi})\mathbf{W}_{\mathrm{G}}) \odot ((\mathbf{R}_{\mathrm{h},k}\bm{\Psi}^{\mathrm{*}}\mathbf{W}_{\mathrm{h},k}) \otimes \mathbbm{1}_{M_{\mathrm{B}}}))^{\mathrm{H}},
\end{align}
where $\mathbbm{1}_{M_{\mathrm{B}}}$ denotes an $M_{\mathrm{B}} \times M_{\mathrm{B}}$ matrix with all elements being $1$ and $\odot$ indicates Hadamard product.

\end{thm}

Hence, the estimation MSE of the effective channel $\hat{\mathbf{H}}_{\mathrm{c},k}, \; k \in \mathcal{K}$, is given as $\mathsf{Tr} \{ \mathbf{C}_{\mathrm{H}_{\mathrm{c},k}} \}, \; k \in \mathcal{K}$, which is actually a function of the pilot sequence $\bm{\Psi}$. In the subsequent exposition, we define $\bm{\Psi} = \bm{\Phi}\mathbf{P}$ with $\bm{\Phi} \triangleq [\bm{\phi}^{(1)}, \cdots, \bm{\phi}^{(T)}]^{\mathrm{T}}$ and $\mathbf{P} \triangleq \mathsf{Diag}(\sqrt{\alpha_{1}}, \cdots, \sqrt{\alpha_{T}})$. Besides, we denote $\bm{\mathcal{W}}_{\mathrm{h}} \triangleq \{ \mathbf{W}_{\mathrm{h},k} \}_{k=1}^{K}$.

Based on the above exposition, we aim at minimizing the overall MSE of all users' effective channels $\{ \hat{\mathbf{H}}_{\mathrm{c},k} \}_{k=1}^{K}$ via designing the pilot sequence and the linear receivers $(\bm{\Phi}, \mathbf{P}, \mathbf{W}_{\mathrm{G}}, \bm{\mathcal{W}}_{\mathrm{h}})$. This task is formulated as the following problem
\begin{align}
(\mathcal{P}1): \min_{\bm{\Phi}, \mathbf{P}, \mathbf{W}_{\mathrm{G}}, \bm{\mathcal{W}}_{\mathrm{h}}} \; &\sum_{k=1}^{K}\frac{\mathsf{Tr} \{ \mathbf{C}_{\mathrm{H}_{\mathrm{c},k}}(\bm{\Phi}, \mathbf{P}, \mathbf{W}_{\mathrm{G}}, \bm{\mathcal{W}}_{\mathrm{h}}) \}}{K} \\
\mathrm{s.t.} \; &\| \bm{\Phi}\mathbf{P} \|_{\mathrm{F}}^{2} \leq P_{\mathrm{max}}, \tag{11a}\\
&|[\bm{\Phi}]_{i,j}| = 1, \; i,j \in \mathcal{M}, \tag{11b}\\
&\alpha_{t} \geq 0, \; t \in \mathcal{T}, \tag{11c}
\end{align}
where $P_{\mathrm{max}}$ in (11a) represents the overall transmission power budget for training. The problem $(\mathcal{P}1)$ is highly challenging due to its nonconvex objective. Especially, it is a nonconvex quartic function in $(\bm{\Phi},\mathbf{P})$. In the following, we propose two algorithms to deal with $(\mathcal{P}1)$.

\section{Pilot Sequence Design}

In this section, we develop efficient solutions to attack the pilot design problem $(\mathcal{P}1)$.

\subsection*{A. GD-Based Solution to $(\mathcal{P}1)$}

In the following, we resolve $(\mathcal{P}1)$ in a block coordinate descent (BCD) [33] manner.

\emph{1) Optimize the phase shift matrix $\bm{\Phi}$}

Fixing other variables, the optimization w.r.t. $\bm{\Phi}$ is given by
\begin{align}
(\mathcal{P}2): \min_{\bm{\Phi}} \; &{\sum}_{k=1}^{K}\mathsf{Tr} \{ \mathbf{C}_{\mathrm{H}_{\mathrm{c},k}}(\bm{\Phi}|\mathbf{P},\mathbf{W}_{\mathrm{G}},\bm{\mathcal{W}}_{\mathrm{h}}) \} \\
\mathrm{s.t.} \; &|[\bm{\Phi}]_{i,j}| = 1, \; i,j \in \mathcal{M}, \tag{12a}
\end{align}
which is non-convex due to the equality constraint (12a) and challenging to solve. Hence, we turn to investigate another optimization equivalent to $(\mathcal{P}2)$ which can be expressed as
\begin{align}
(\mathcal{P}3): \min_{\bm{\Theta}} \; &{\sum}_{k=1}^{K}\mathsf{Tr} \{ \mathbf{C}_{\mathrm{H}_{\mathrm{c},k}}(\bm{\Theta}|\mathbf{P},\mathbf{W}_{\mathrm{G}},\bm{\mathcal{W}}_{\mathrm{h}}) \} \\
\mathrm{s.t.} \; &[\bm{\Theta}]_{i,j} \in [0,2\pi], \; i,j \in \mathcal{M}, \tag{13a}
\end{align}
where $\bm{\Theta} \triangleq \angle(\bm{\Phi})$ and $\angle(\bm{\Phi})$ takes the value of phases of $\bm{\Phi}$'s entries in an element-wise manner. By noting that $e^{j(\theta + 2n\pi)} = e^{j\theta}, \; n \in \mathbb{Z}$, problem $(\mathcal{P}3)$ can be viewed as an unconstrained optimization problem w.r.t. $\bm{\Theta}$. Therefore, we adopt GD algorithm to tackle $(\mathcal{P}3)$. Denote $\mathsf{g}(\bm{\Theta})$ as the objective of $(\mathcal{P}3)$. Then, the gradient descent of $\bm{\Theta}$ is conducted as
\begin{equation}
\bm{\Theta}^{(n + 1)} := \bm{\Theta}^{(n)} - \eta_{\bm{\Theta}}^{(n)}\nabla_{\bm{\Theta}}\mathsf{g}(\bm{\Theta})|_{\bm{\Theta} = \bm{\Theta}^{(n)}},
\end{equation}
where $\eta_{\bm{\Theta}}^{(n)}$ is the step size for updating $\bm{\Theta}$. Generally, constant or diminishing step size rule can yield satisfactory convergence [34]. We relegate the derivation of $\nabla_{\bm{\Theta}}\mathsf{g}(\bm{\Theta})$ to Appendix B. Lastly, $\bm{\Phi}^{(n + 1)}$ is given by $\bm{\Phi}^{(n + 1)} = e^{j\bm{\Theta}^{(n + 1)}}$.

\emph{2) Optimize the power allocation $\mathbf{P}$}

We proceed to study the optimization of power allocation $\mathbf{P}$ when other variables are given. Notice again $\mathbf{P} \triangleq \mathsf{Diag}(\sqrt{\alpha_{1}}, \cdots, \sqrt{\alpha_{T}})$. Therefore, by defining $\mathbf{p} \triangleq \mathsf{diag}(\mathbf{P})$, the sub-problem to optimize power allocation can be expressed as
\begin{align}
(\mathcal{P}4): \min_{\mathbf{p} \in \mathbb{R}^{T}} \; &{\sum}_{k=1}^{K}\mathsf{Tr} \{ \mathbf{C}_{\mathrm{H}_{\mathrm{c},k}}(\mathbf{p}|\bm{\Phi},\mathbf{W}_{\mathrm{G}},\bm{\mathcal{W}}_{\mathrm{h}}) \} \\
\mathrm{s.t.} \; &\mathbf{p}^{\mathrm{T}}\mathbf{p} \leq \frac{P_{\mathrm{max}}}{M_{\mathrm{R}}}, \tag{15a}\\
&\mathbf{p} \geq \bm{0}. \tag{15b}
\end{align}

The objective of $(\mathcal{P}4)$ is non-convex and we still adopt GD-like method to attack it. Note that plain GD algorithm only applies to unconstrained optimization problem. For the constrained problem $(\mathcal{P}4)$, we adopt the gradient projection (GP) method [34]. GP method is an iterative procedure. In each iteration, it performs two steps in order: i) gradient descent and ii) projection onto the feasible domain. Denoting the objective of $(\mathcal{P}4)$ as $\mathsf{h}(\mathbf{p})$, we elaborate these two steps in the following:

\begin{itemize}

\item Gradient Descent: The variable $\mathbf{p}$ moves in the direction of negative derivative, i.e.
\begin{equation}
\mathbf{p}^{(n + 1)} := \mathbf{p}^{(n)} - \eta_{\mathrm{p}}^{(n)}\nabla_{\mathbf{p}}\mathsf{h}(\mathbf{p})|_{\mathbf{p} = \mathbf{p}^{(n)}},
\end{equation}
where $\eta_{\mathrm{p}}^{(n)}$ is the step size for updating power allocation, which can be set by following the step size rules presented in [34]. The calculation of $\nabla_{\mathbf{p}}\mathsf{h}(\mathbf{p})$ is shown in Appendix C.

\item Projection: After performing the update of $\mathbf{p}$ as shown above, there exists a risk that $\mathbf{p}^{(n + 1)}$ moves out of the feasible domain identified by (15a) and (15b). Hence, we need project $\mathbf{p}^{(n + 1)}$ onto the convex feasible domain of $(\mathcal{P}4)$ [34], which means to solve the following optimization
\begin{align}
(\mathcal{P}5): \min_{\mathbf{x} \in \mathbb{R}^{T}} \; &\frac{1}{2}\| \mathbf{x} - \mathbf{p}^{(n + 1)} \|_{2}^{2} \\
\mathrm{s.t.} \; &\mathbf{x}^{\mathrm{T}}\mathbf{x} \leq \frac{P_{\mathrm{max}}}{M_{\mathrm{R}}}, \tag{17a}\\
&\mathbf{x} \geq \bm{0}. \tag{17b}
\end{align}

The optimal solution to $(\mathcal{P}5)$ is given by the following Theorem 2, whose proof is left to Appendix D.
\begin{thm}

If $\| [\mathbf{p}^{(n + 1)}]_{+} \|_{2}^{2} \leq \frac{P_{\mathrm{max}}}{M_{\mathrm{R}}}$, we have
\begin{equation}
\mathbf{x}^{\star} = [\mathbf{p}^{(n + 1)}]_{+},
\end{equation}
otherwise,
\begin{equation}
\mathbf{x}^{\star} = \sqrt{\frac{P_{\mathrm{max}}}{M_{\mathrm{R}}}}\frac{[\mathbf{p}^{(n + 1)}]_{+}}{\| [\mathbf{p}^{(n + 1)}]_{+} \|_{2}},
\end{equation}
where $[\mathbf{p}^{(n + 1)}]_{+} \triangleq \max \{ \mathbf{p}^{(n + 1)}, \bm{0} \}$ in an element-wise manner.

\end{thm}

\end{itemize}

\emph{3) Optimize the linear receiver at the BS $\mathbf{W}_{\mathrm{G}}$}

When other variables are given, the subproblem w.r.t. $\mathbf{W}_{\mathrm{G}}$ is indeed an unconstrained convex quadratic optimization problem. Therefore, by checking the first-order optimality condition of $\mathbf{W}_{\mathrm{G}}$, the newly obtained $\mathbf{W}_{\mathrm{G}}$ can be expressed in the closed-form as follows
\begin{equation}
\mathbf{W}_{\mathrm{G}} = \mathbf{J}_{\mathrm{G}}\mathbf{D}_{\mathrm{G},1}\mathbf{D}_{\mathrm{G},2}^{-1},
\end{equation}
where
\begin{align}
&\mathbf{J}_{\mathrm{G}} = (\mathsf{A}_{\mathrm{G}}(\bm{\Psi})\mathbf{R}_{\mathrm{G}}\mathsf{A}_{\mathrm{G}}^{\mathrm{H}}(\bm{\Psi}) + \bm{\Sigma}_{\mathrm{B}} \otimes \mathbf{I}_{M_{\mathrm{R}}})^{-1}\mathsf{A}_{\mathrm{G}}(\bm{\Psi})\mathbf{R}_{\mathrm{G}}, \notag\\
&\mathbf{D}_{\mathrm{G},1} = {\sum}_{k=1}^{K}\mathsf{Ddiag}((\mathbf{W}_{\mathrm{h},k}^{\mathrm{H}}\bm{\Psi}^{\mathrm{T}}\mathbf{R}_{\mathrm{h},k}) \otimes \mathbbm{1}_{M_{\mathrm{B}}}), \notag\\
&\mathbf{D}_{\mathrm{G},2} = {\sum}_{k=1}^{K}\mathsf{Ddiag}(\mathbf{C}_{\hat{\mathrm{h}},k} \otimes \mathbbm{1}_{M_{\mathrm{B}}}),
\end{align}
with $\mathsf{Ddiag}(\mathbf{X})$ constructing a diagonal matrix with its diagonal elements being those of the square matrix $\mathbf{X}$. The derivation of (20) is detailed in Appendix E.

\emph{4) Optimize the linear receiver at every user $\bm{\mathcal{W}}_{\mathrm{h}}$}

Following the similar procedure of updating $\mathbf{W}_{\mathrm{G}}$, the solution to $\mathbf{W}_{\mathrm{h},k}, \; k \in \mathcal{K}$, can also be obtained in a closed-form as follows (details are omitted for brevity)
\begin{equation}
\mathbf{W}_{\mathrm{h},k} = \mathbf{J}_{\mathrm{h},k}\mathbf{D}_{\mathrm{h},1}\mathbf{D}_{\mathrm{h},2}^{-1}, \; k \in \mathcal{K},
\end{equation}
where
\begin{equation}
\mathbf{J}_{\mathrm{h},k} = (\bm{\Psi}^{\mathrm{T}}\mathbf{R}_{\mathrm{h},k}\bm{\Psi}^{*} + \bm{\Sigma}_{\mathrm{U},k})^{-1}\bm{\Psi}^{\mathrm{T}}\mathbf{R}_{\mathrm{h},k}, \; k \in \mathcal{K},
\end{equation}
$\mathbf{D}_{\mathrm{h},1}$ and $\mathbf{D}_{\mathrm{h},2}$ are $M_{\mathrm{R}}$-dimensional diagonal matrices with $[\mathbf{D}_{\mathrm{h},1}]_{m,m} = \sum_{n=1}^{M_{\mathrm{B}}}[\mathbf{W}_{\mathrm{G}}^{\mathrm{H}}\mathsf{A}_{\mathrm{G}}(\bm{\Psi})\mathbf{R}_{\mathrm{G}}]_{(m - 1)M_{\mathrm{B}} + n,(m - 1)M_{\mathrm{B}} + n}, \; m \in \mathcal{M}$, and $[\mathbf{D}_{\mathrm{h},2}]_{m,m} = \sum_{n=1}^{M_{\mathrm{B}}}[\mathbf{C}_{\hat{\mathrm{G}}}]_{(m - 1)M_{\mathrm{B}} + n,(m - 1)M_{\mathrm{B}} + n}, \; m \in \mathcal{M}$, respectively.

The overall algorithm of solving $(\mathcal{P}1)$ based on GD is summarized in Algorithm 1.

\begin{algorithm}[!t]
\caption{Solving the problem $(\mathcal{P}1)$ based on GD}
\begin{algorithmic}[1]
\STATE Initialize feasible $\bm{\Psi}^{(0)} = \bm{\Phi}^{(0)}\mathbf{P}^{(0)}$ and $t = 0$;
\REPEAT
\STATE update $\mathbf{W}_{\mathrm{G}}^{(t + 1)}$ by (20);
\STATE update $\bm{\mathcal{W}}_{\mathrm{h}}^{(t + 1)}$ by (22);
\STATE update $\bm{\Phi}^{(t + 1)}$ by solving $(\mathcal{P}2)$ via GD procedure;
\STATE update $\mathbf{P}^{(t + 1)}$ by solving $(\mathcal{P}4)$ via GP procedure;
\STATE set $\bm{\Psi}^{(t + 1)} = \bm{\Phi}^{(t + 1)}\mathbf{P}^{(t + 1)}$;
\STATE $t := t + 1$;
\UNTIL{\textcolor{blue}{a certain stopping criterion is reached}}
\end{algorithmic}
\end{algorithm}

\subsection*{B. PDD-Based Solution to $(\mathcal{P}1)$}

One potential defect of the previously discussed GD-based solution is its onerous gradient evaluation procedures, as reflected in Appendix B $\&$ C. In this subsection, we adopt the PDD methodology [27] to propose an alternative solution that updates all block coordinates highly efficiently. To this end, we first transform the difficult objective in (11) into a more tractable form amiable to block coordinate update via introducing auxiliary variables. Specifically, we introduce multiple copies of $\bm{\Phi}\mathbf{P}$ by defining
\begin{align}
&\mathbf{U}_{1,k} = (\bm{\Phi}\mathbf{P})^{*}, \; k \in \mathcal{K},\\
&\mathbf{U}_{2} = (\bm{\Phi}\mathbf{P})^{*} \otimes \mathbf{I}_{M_{\mathrm{B}}},
\end{align}
and rewrite the original complicated objective into simple forms of the newly introduced copies as follows
\begin{align}
&\mathbf{U}_{3,k} = \mathbf{R}_{\mathrm{h},k}^{\frac{1}{2}}\mathbf{U}_{1,k}\mathbf{W}_{\mathrm{h},k}, \; k \in \mathcal{K},\\
&\mathbf{U}_{4} = \mathbf{R}_{\mathrm{G}}^{\frac{1}{2}}\mathbf{U}_{2}\mathbf{W}_{\mathrm{G}},\\
&\mathbf{F}_{1,k}(\mathbf{U}_{3,k},\mathbf{W}_{\mathrm{h},k}) \!\triangleq\! \tilde{\mathbf{F}}_{1,k}(\mathbf{U}_{3,k},\mathbf{W}_{\mathrm{h},k}) \otimes \mathbbm{1}_{M_{\mathrm{B}}}, \; k \in \mathcal{K},\\
&\mathbf{F}_{2}(\mathbf{U}_{4},\mathbf{W}_{\mathrm{G}}) \triangleq \mathbf{U}_{4}^{\mathrm{H}}\mathbf{U}_{4} + \mathbf{W}_{\mathrm{G}}^{\mathrm{H}}(\bm{\Sigma}_{\mathrm{B}} \otimes \mathbf{I}_{M_{\mathrm{R}}})\mathbf{W}_{\mathrm{G}},\\
&\tilde{\mathbf{F}}_{1,k}(\mathbf{U}_{3,k},\!\mathbf{W}_{\mathrm{h},k}) \!\triangleq\! \mathbf{U}_{3,k}^{\mathrm{H}}\mathbf{U}_{3,k} \!\!+\!\! \mathbf{W}_{\mathrm{h},k}^{\mathrm{H}}\bm{\Sigma}_{\mathrm{U},k}\mathbf{W}_{\mathrm{h},k}, k \!\in\! \mathcal{K}.
\end{align}
Therefore, the problem $(\mathcal{P}1)$ can be equivalently rewritten as
\begin{align}
&(\mathcal{P}6): \min_{\bm{\mathcal{V}}} \; \sum_{k=1}^{K}\mathsf{Tr} \{ \mathbf{F}_{2}(\mathbf{U}_{4},\mathbf{W}_{\mathrm{G}}) \!\odot\! \mathbf{F}_{1,k}(\mathbf{U}_{3,k},\mathbf{W}_{\mathrm{h},k}) \!+\! \mathbf{R}_{\mathrm{G}}\notag\\
&\!\odot\! (\mathbf{R}_{\mathrm{h},k} \!\!\otimes\!\! \mathbbm{1}_{M_{\mathrm{B}}}) \!\!-\!\! 2\mathsf{Re} \{ (\mathbf{R}_{\mathrm{G}}^{\frac{1}{2}}\mathbf{U}_{4}) \!\odot\! ((\mathbf{R}_{\mathrm{h},k}^{\frac{1}{2}}\mathbf{U}_{3,k}) \!\otimes\! \mathbbm{1}_{M_{\mathrm{B}}}) \} \} \\
&\qquad\quad\mathrm{s.t.} \; \mathrm{(11a)-(11c)}, \tag{31a}\\
&\qquad\qquad\quad\mathrm{(24)-(29)}, \tag{31b}
\end{align}
where the newly introduced notations above are defined as
\begin{align}
&\bm{\mathcal{V}} \triangleq \{ \bm{\mathcal{U}}_{1}, \mathbf{U}_{2}, \bm{\mathcal{U}}_{3}, \mathbf{U}_{4}, \bm{\Phi}, \mathbf{P}, \bm{\mathcal{W}}_{\mathrm{h}}, \mathbf{W}_{\mathrm{G}} \}, \notag\\
&\bm{\mathcal{U}}_{1} \triangleq \{ \mathbf{U}_{1,k} \}_{k=1}^{K}, \; \bm{\mathcal{U}}_{3} \triangleq \{ \mathbf{U}_{3,k} \}_{k=1}^{K}.
\end{align}

Following the PDD method [27], to tackle $(\mathcal{P}6)$, we turn to solve its augmented Lagrangian problem given as follows

\vspace{-4mm}
{\small
\begin{align}
&(\mathcal{P}7): \min_{\bm{\mathcal{V}}} \; \mathsf{f}_{\rho}(\bm{\mathcal{V}}) \!+\! \!\sum_{k=1}^{K}\!\mathsf{Tr} \{ \mathbf{F}_{2}(\mathbf{U}_{4},\mathbf{W}_{\mathrm{G}}) \!\odot\! \mathbf{F}_{1,k}(\mathbf{U}_{3,k},\mathbf{W}_{\mathrm{h},k}) \!+\! \mathbf{R}_{\mathrm{G}}\notag\\
&\!\odot\! (\mathbf{R}_{\mathrm{h},k} \!\!\otimes\!\! \mathbbm{1}_{M_{\mathrm{B}}}) \!\!-\!\! 2\mathsf{Re} \{ (\mathbf{R}_{\mathrm{G}}^{\frac{1}{2}}\mathbf{U}_{4}) \!\odot\! ((\mathbf{R}_{\mathrm{h},k}^{\frac{1}{2}}\mathbf{U}_{3,k}) \!\otimes\! \mathbbm{1}_{M_{\mathrm{B}}}) \} \} \\
&\qquad\quad\mathrm{s.t.} \; \mathrm{(11a)-(11c)}, \tag{33a}
\end{align}}

\vspace{-5mm}
\noindent with the augmented term $\mathsf{f}_{\rho}(\bm{\mathcal{V}})$ defined as
\begin{align}
&\mathsf{f}_{\rho}(\bm{\mathcal{V}}) \triangleq \frac{1}{2\rho}\bigg(\sum_{k=1}^{K}\|\mathbf{U}_{3,k} - \mathbf{R}_{\mathrm{h},k}^{\frac{1}{2}}\mathbf{U}_{1,k}\mathbf{W}_{\mathrm{h},k} + \rho\bm{\Lambda}_{3,k}\|_{\mathrm{F}}^{2}\notag\\
&+ \!\!\sum_{k=1}^{K}\!\|\mathbf{U}_{1,k} \!\!-\!\! (\bm{\Phi}\mathbf{P})^{*} \!\!+\!\! \rho\bm{\Lambda}_{1,k}\|_{\mathrm{F}}^{2} \!+\! \|\mathbf{U}_{2} \!\!-\!\! (\bm{\Phi}\mathbf{P})^{*} \!\otimes\! \mathbf{I}_{M_{\mathrm{B}}} \!\!+\!\! \rho\bm{\Lambda}_{2}\|_{\mathrm{F}}^{2}\notag\\
&+ \|\mathbf{U}_{4} - \mathbf{R}_{\mathrm{G}}^{\frac{1}{2}}\mathbf{U}_{2}\mathbf{W}_{\mathrm{G}} + \rho\bm{\Lambda}_{4}\|_{\mathrm{F}}^{2}\bigg),
\end{align}
where $\rho$ represents the penalty coefficient, $\{ \bm{\Lambda}_{1,k} \}_{k=1}^{K}$, $\bm{\Lambda}_{2}$, $\{ \bm{\Lambda}_{3,k} \}_{k=1}^{K}$ and $\bm{\Lambda}_{4}$ represent the Lagrangian multipliers associated with the constraints (24)-(27), respectively.

The PDD method is a two-layer iterative procedure [27], with its inner layer alternatively optimizing the different blocks of variables comprising $\bm{\mathcal{V}}$ and its outer layer selectively updating the dual variables $\bm{\Lambda}_{i}$, $i = 1, 2, 3, 4$ or the penalty coefficient $\rho$. The two layers' updating procedure will be specified in the sequel.

For the inner layer, we adopt the block successive upper-bound minimization (BSUM) to update various blocks [27], [35], with each block's update being elaborated as follows.

\emph{1) Optimize the phase shift matrix $\bm{\Phi}$}

When other variables are given, the subproblem w.r.t. $\bm{\Phi}$ can be expressed as
\begin{align}
(\mathcal{P}8): \min_{\bm{\Phi}} \; &{\sum}_{k=1}^{K}\| \mathbf{U}_{1,k} - (\bm{\Phi}\mathbf{P})^{*} + \rho\bm{\Lambda}_{1,k} \|_{\mathrm{F}}^{2}\notag\\
&+ \| \mathbf{U}_{2} - (\bm{\Phi}\mathbf{P})^{*} \otimes \mathbf{I}_{M_{\mathrm{B}}} + \rho\bm{\Lambda}_{2} \|_{\mathrm{F}}^{2}\\
\mathrm{s.t.} \; &|[\bm{\Phi}]_{i,j}| = 1, \; i,j \in \mathcal{M}. \tag{35a}
\end{align}
Note that the quadratic term $\| \bm{\Phi}\mathbf{P} \|_{\mathrm{F}}^{2}$ reduces to a constant $T\sum_{t=1}^{T}\alpha_{t}$ due to (35a), which is independent of $\bm{\Phi}$. Hence, utilizing the above fact, the problem $(\mathcal{P}8)$ boils down to maximizing a linear objective given as follows
\begin{align}
(\mathcal{P}9): \max_{\bm{\Phi}} \; &\mathsf{Re} \{ \mathsf{Tr} \{ \mathbf{Q}\bm{\Phi} \} \} \\
\mathrm{s.t.} \; &|[\bm{\Phi}]_{i,j}| = 1, \; i,j \in \mathcal{M}, \tag{36a}
\end{align}
where the parameter $\mathbf{Q}$ is defined as
\begin{equation}
\mathbf{Q} \triangleq {\sum}_{k=1}^{K}\mathbf{P}(\mathbf{U}_{1,k}^{\mathrm{T}} + \rho\bm{\Lambda}_{1,k}^{\mathrm{T}}) + \tilde{\mathbf{C}}_{\bm{\Phi}},
\end{equation}
and the $(i,j)$th element of the matrix $\tilde{\mathbf{C}}_{\bm{\Phi}}$ is given by $\sum_{m=1}^{M_{\mathrm{B}}}[(\mathbf{P} \!\otimes\! \mathbf{I}_{M_{\mathrm{B}}})(\mathbf{U}_{2}^{\mathrm{T}} \!+\! \rho\bm{\Lambda}_{2}^{\mathrm{T}})]_{(i \!-\! 1)M_{\mathrm{B}} \!+\! m,(j \!-\! 1)M_{\mathrm{B}} \!+\! m}, \; i,j \in \mathcal{M}$.

The problem $(\mathcal{P}9)$, although nonconvex, achieves optimum when the elements' phases of $\bm{\Phi}$ align with those of $\mathbf{Q}^{\mathrm{H}}$. Consequently, the optimal solution of $(\mathcal{P}8)$ is given analytically by
\begin{equation}
\bm{\Phi} = e^{j\angle(\mathbf{Q}^{\mathrm{H}})}.
\end{equation}

\emph{2) Optimize the power allocation $\mathbf{P}$}

After some manipulations, the update of $\mathbf{P}$ with the remaining variables fixed can be written as
\begin{align}
(\mathcal{P}10): \min_{\mathbf{p} \in \mathbb{R}^{T}} \; &(KM_{\mathrm{R}} + M_{\mathrm{B}}M_{\mathrm{R}})\mathbf{p}^{\mathrm{T}}\mathbf{p} - 2\mathsf{Re} \{ \mathbf{q}^{\mathrm{H}}\mathbf{p} \} \\
\mathrm{s.t.} \; &\mathbf{p}^{\mathrm{T}}\mathbf{p} \leq \frac{P_{\mathrm{max}}}{M_{\mathrm{R}}}, \tag{39a}\\
&\mathbf{p} \geq \bm{0}, \tag{39b}
\end{align}
where $\mathbf{p} \!\triangleq\! \mathsf{diag}({\mathbf{P}})$, $\mathbf{q}^{\mathrm{H}} \triangleq \big(\mathsf{diag}\big(\sum_{k=1}^{K}(\mathbf{U}_{1,k}^{\mathrm{T}} + \rho\bm{\Lambda}_{1,k}^{\mathrm{T}})\bm{\Phi}\big)\big)^{\mathrm{T}} + \mathbf{c}_{2}^{\mathrm{T}}$ and $[\mathbf{c}_{2}]_{m} = \sum_{n=1}^{M_{\mathrm{B}}}[(\mathbf{U}_{2}^{\mathrm{T}} + \rho\bm{\Lambda}_{2}^{\mathrm{T}})(\bm{\Phi} \otimes \mathbf{I}_{M_{\mathrm{B}}})]_{(m - 1)M_{\mathrm{B}} + n,(m - 1)M_{\mathrm{B}} + n}, \; m \in \mathcal{M}$.

Following the similar procedure as solving $(\mathcal{P}5)$, the optimal solution to $(\mathcal{P}10)$ is given by
\begin{itemize}

\item if $\big\| \frac{[\mathsf{Re} \{ \mathbf{q} \}]_{+}}{KM_{\mathrm{R}} + M_{\mathrm{B}}M_{\mathrm{R}}} \big\|_{2}^{2} \leq \frac{P_{\mathrm{max}}}{M_{\mathrm{R}}}$, we have
\begin{equation}
\mathbf{p}^{\star} = \frac{[\mathsf{Re} \{ \mathbf{q} \}]_{+}}{KM_{\mathrm{R}} + M_{\mathrm{B}}M_{\mathrm{R}}},
\end{equation}

\item otherwise,

\begin{equation}
\mathbf{p}^{\star} = \sqrt{\frac{P_{\mathrm{max}}}{M_{\mathrm{R}}}}\frac{[\mathsf{Re} \{ \mathbf{q} \}]_{+}}{\| [\mathsf{Re} \{ \mathbf{q} \}]_{+} \|_{2}},
\end{equation}

\end{itemize}
which can be proven by exploiting Appendix D.

\emph{3) Optimize the auxiliary variable $\bm{\mathcal{U}}_{1}$}

Obviously, the optimization w.r.t. $\bm{\mathcal{U}}_{1}$ is an unconstrained convex quadratic minimization problem. Hence, by setting the first-order derivative of (33) w.r.t. $\mathbf{U}_{1,k}, \; k \in \mathcal{K}$, to zero, we attain the following equation
\begin{equation}
\mathbf{R}_{\mathrm{h},k}^{-1}\mathbf{U}_{1,k} + \mathbf{U}_{1,k}\mathbf{W}_{\mathrm{h},k}\mathbf{W}_{\mathrm{h},k}^{\mathrm{H}} = \tilde{\mathbf{C}}_{1,k}, \; k \in \mathcal{K},
\end{equation}
where
\begin{equation}
\tilde{\mathbf{C}}_{1,k} \!\triangleq\! \mathbf{R}_{\mathrm{h},k}^{-1}((\bm{\Phi}\mathbf{P})^{*} \!-\! \rho\bm{\Lambda}_{1,k} \!+\! \mathbf{R}_{\mathrm{h},k}^{\frac{1}{2}}(\mathbf{U}_{3,k} \!+\! \rho\bm{\Lambda}_{3,k})\mathbf{W}_{\mathrm{h},k}^{\mathrm{H}}).
\end{equation}
The linear system in (42) is the well-known Sylvester equation and can be readily solved via the seminal Hessenburg-Schur algorithm, which has the complexity of $\mathcal{O}(M_{\mathrm{R}}^{3})$ [36].$^{2}$\footnote{$^{2}$Note in MATLAB, the standard build-in function $\mathsf{sylvester}()$ can be invoked to solve (42) immediately.}

\emph{4) Optimize the auxiliary variable $\mathbf{U}_{2}$}

Following the similar arguments as in updating $\bm{\mathcal{U}}_{1}$, the optimization of $\mathbf{U}_{2}$ also reduces to solving a Sylvester equation given as follows
\begin{equation}
\mathbf{R}_{\mathrm{G}}^{-1}\mathbf{U}_{2} + \mathbf{U}_{2}\mathbf{W}_{\mathrm{G}}\mathbf{W}_{\mathrm{G}}^{\mathrm{H}} = \tilde{\mathbf{C}}_{2},
\end{equation}
where
\begin{equation}
\tilde{\mathbf{C}}_{2} \triangleq \mathbf{R}_{\mathrm{G}}^{-1}(\mathsf{A}_{\mathrm{G}}^{\mathrm{H}}(\bm{\Psi}) - \rho\bm{\Lambda}_{2} + \mathbf{R}_{\mathrm{G}}^{\frac{1}{2}}(\mathbf{U}_{4} + \rho\bm{\Lambda}_{4})\mathbf{W}_{\mathrm{G}}^{\mathrm{H}}).
\end{equation}

\emph{5) Optimize the auxiliary variable $\bm{\mathcal{U}}_{3}$}

Next, we proceed to study the update of $\bm{\mathcal{U}}_{3}$, which is an unconstrained convex quadratic problem, whose closed-form solution can be obtained via checking the first-order optimality condition, shown as
\begin{align}
\mathbf{U}_{3,k} = &\bigg(\frac{1}{2\rho}\mathbf{R}_{\mathrm{h},k}^{\frac{1}{2}}\mathbf{U}_{1,k}\mathbf{W}_{\mathrm{h},k} - \frac{1}{2}\bm{\Lambda}_{3,k} + \mathbf{R}_{\mathrm{h},k}^{\frac{1}{2}}\tilde{\mathbf{C}}_{\mathrm{d},1}\bigg) \notag\\
&\times \bigg(\frac{1}{2\rho}\mathbf{I}_{M_{\mathrm{R}}} + \tilde{\mathbf{C}}_{\mathrm{d},2}\bigg)^{-1}, \; k \in \mathcal{K},
\end{align}
where $\tilde{\mathbf{C}}_{\mathrm{d},1}$ and $\tilde{\mathbf{C}}_{\mathrm{d},2}$ are $M_{\mathrm{R}}$-dimensional diagonal matrices whose $i$th diagonal elements, $i \in \mathcal{M}$, equal to $\sum_{j=(i - 1)M_{\mathrm{B}} + 1}^{iM_{\mathrm{B}}}[\mathbf{U}_{4}^{\mathrm{H}}\mathbf{R}_{\mathrm{G}}^{\frac{1}{2}}]_{jj}$ and $\sum_{j=(i - 1)M_{\mathrm{B}} + 1}^{iM_{\mathrm{B}}}[\mathbf{U}_{4}^{\mathrm{H}}\mathbf{U}_{4} + \mathbf{W}_{\mathrm{G}}^{\mathrm{H}}(\bm{\Sigma}_{\mathrm{B}} \otimes \mathbf{I}_{M_{\mathrm{R}}})\mathbf{W}_{\mathrm{G}}]_{jj}$, respectively.

\emph{6) Optimize the auxiliary variable $\mathbf{U}_{4}$}

Similar to the optimization of $\bm{\mathcal{U}}_{3}$, the solution to $\mathbf{U}_{4}$ can be expressed in the closed-form as
\begin{align}
\mathbf{U}_{4} = &\bigg(\frac{1}{2\rho}\mathbf{R}_{\mathrm{G}}^{\frac{1}{2}}\mathbf{U}_{2}\mathbf{W}_{\mathrm{G}} - \frac{1}{2}\bm{\Lambda}_{4} + \sum_{k=1}^{K}\mathbf{R}_{\mathrm{G}}^{\frac{1}{2}}\mathsf{Ddiag}((\mathbf{U}_{3,k}^{\mathrm{H}}\mathbf{R}_{\mathrm{h},k}^{\frac{1}{2}}) \notag\\
&\otimes \mathbbm{1}_{M_{\mathrm{B}}})\bigg)\bigg(\frac{1}{2\rho}\mathbf{I}_{M_{\mathrm{B}}M_{\mathrm{R}}} + \tilde{\mathbf{C}}_{\mathrm{d},3}\bigg)^{-1},
\end{align}
where
\begin{equation}
\tilde{\mathbf{C}}_{\mathrm{d},3} \!\triangleq\! \sum_{k=1}^{K}\mathsf{Ddiag}((\mathbf{U}_{3,k}^{\mathrm{H}}\mathbf{U}_{3,k} \!+\! \mathbf{W}_{\mathrm{h},k}^{\mathrm{H}}\bm{\Sigma}_{\mathrm{U},k}\mathbf{W}_{\mathrm{h},k}) \!\otimes\! \mathbbm{1}_{M_{\mathrm{B}}}).
\end{equation}

\emph{7) Optimize the linear receiver at every user $\bm{\mathcal{W}}_{\mathrm{h}}$}

The update of $\bm{\mathcal{W}}_{\mathrm{h}}$ is an unconstrained convex quadratic optimization problem, whose optimal solution is equivalent to solving the following Sylvester equation
\begin{equation}
\frac{1}{2\rho}\bm{\Sigma}_{\mathrm{U},k}^{-1}\mathbf{U}_{1,k}^{\mathrm{H}}\mathbf{R}_{\mathrm{h},k}\mathbf{U}_{1,k}\mathbf{W}_{\mathrm{h},k} \!\!+\!\! \mathbf{W}_{\mathrm{h},k}\tilde{\mathbf{C}}_{\mathrm{d},2} \!\!=\!\! \tilde{\mathbf{C}}_{\mathrm{W_{h}},k}, k \!\in\! \mathcal{K},
\end{equation}
where
\begin{equation}
\tilde{\mathbf{C}}_{\mathrm{W_{h}},k} \triangleq \frac{1}{2\rho}\bm{\Sigma}_{\mathrm{U},k}^{-1}\mathbf{U}_{1,k}^{\mathrm{H}}\mathbf{R}_{\mathrm{h},k}^{\frac{1}{2}}(\mathbf{U}_{3,k} + \rho\bm{\Lambda}_{3,k}), \; k \in \mathcal{K}.
\end{equation}

\emph{8) Optimize the linear receiver at the BS $\mathbf{W}_{\mathrm{G}}$}

The update of $\mathbf{W}_{\mathrm{G}}$ is similar to that of $\bm{\mathcal{W}}_{\mathrm{h}}$ and can be conducted by solving the following Sylvester equation (details are skipped to avoid repetition)
\begin{align}
&\frac{1}{2\rho}(\bm{\Sigma}_{\mathrm{B}} \otimes \mathbf{I}_{M_{\mathrm{R}}})^{-1}\mathbf{U}_{2}^{\mathrm{H}}\mathbf{R}_{\mathrm{G}}\mathbf{U}_{2}\mathbf{W}_{\mathrm{G}} + \mathbf{W}_{\mathrm{G}}\tilde{\mathbf{C}}_{\mathrm{d},3} \notag\\
&= \frac{1}{2\rho}(\bm{\Sigma}_{\mathrm{B}} \otimes \mathbf{I}_{M_{\mathrm{R}}})^{-1}\mathbf{U}_{2}^{\mathrm{H}}\mathbf{R}_{\mathrm{G}}^{\frac{1}{2}}(\mathbf{U}_{4} + \rho\bm{\Lambda}_{4}).
\end{align}

For the outer layer, if all the equalities (24)-(27) are approximately satisfied, we will update the Lagrangian multipliers in a gradient ascent manner, i.e., $\bm{\Lambda}_{i}^{(t + 1)} := \bm{\Lambda}_{i}^{(t)} + \rho^{-1}(\mathbf{U}_{i} - \mathbf{U}_{i}^{\mathrm{R}})$, where $\mathbf{U}_{i}^{\mathrm{R}}$ represents the right hand side of the $i$th equality. Otherwise, the penalty coefficient $\rho$ should be decreased to force the equalities to be approached in the subsequent iterations.

The proposed PDD-based algorithm of handling $(\mathcal{P}1)$ is stated in Algorithm 2.

\begin{algorithm}[!t]
\caption{Solving the problem $(\mathcal{P}1)$ based on PDD}
\begin{algorithmic}[1]
\STATE Initialize feasible $\bm{\Psi}^{(0)} = \bm{\Phi}^{(0)}\mathbf{P}^{(0)}$ and $t = 0$;
\REPEAT
\STATE set $\bm{\Phi}^{(t,0)} := \bm{\Phi}^{(t)}$, $\mathbf{P}^{(t,0)} := \mathbf{P}^{(t)}$, $\bm{\mathcal{W}}_{\mathrm{h}}^{(t,0)} := \bm{\mathcal{W}}_{\mathrm{h}}^{(t)}$, $\mathbf{W}_{\mathrm{G}}^{(t,0)} := \mathbf{W}_{\mathrm{G}}^{(t)}$, $\{ \mathbf{U}_{i}^{(t,0)} \}_{i=1}^{4} := \{ \mathbf{U}_{i}^{(t)} \}_{i=1}^{4}$ and \\ $s = 0$;
\REPEAT
\STATE update $\bm{\mathcal{W}}_{\mathrm{h}}^{(t,s + 1)}$ by solving equation (49);
\STATE update $\mathbf{W}_{\mathrm{G}}^{(t,s + 1)}$ by solving equation (51);
\STATE update $\bm{\mathcal{U}}_{1}^{(t,s + 1)}$ by solving equation (42);
\STATE update $\mathbf{U}_{2}^{(t,s + 1)}$ by solving equation (44);
\STATE update $\bm{\mathcal{U}}_{3}^{(t,s + 1)}$ by (46);
\STATE update $\mathbf{U}_{4}^{(t,s + 1)}$ by (47);
\STATE update $\bm{\Phi}^{(t,s + 1)}$ by (38);
\STATE update $\mathbf{P}^{(t,s + 1)}$ by solving $(\mathcal{P}10)$;
\STATE $s := s + 1$;
\UNTIL{\textcolor{blue}{a certain stopping criterion is reached}}
\STATE set $\bm{\Phi}^{(t + 1)} := \bm{\Phi}^{(t,\infty)}$, $\mathbf{P}^{(t + 1)} := \mathbf{P}^{(t,\infty)}$, $\bm{\mathcal{W}}_{\mathrm{h}}^{(t + 1)} := \bm{\mathcal{W}}_{\mathrm{h}}^{(t,\infty)}$, $\mathbf{W}_{\mathrm{G}}^{(t + 1)} := \mathbf{W}_{\mathrm{G}}^{(t,\infty)}$ and $\{ \mathbf{U}_{i}^{(t + 1)} \}_{i=1}^{4} := \{ \mathbf{U}_{i}^{(t,\infty)} \}_{i=1}^{4}$;
\STATE set $\bm{\Psi}^{(t + 1)} = \bm{\Phi}^{(t + 1)}\mathbf{P}^{(t + 1)}$;
\FOR{$i = 1$ to $4$}
\IF{$\| \mathbf{U}_{i}^{(t + 1)} - (\mathbf{U}_{i}^{\mathrm{R}})^{(t + 1)} \|_{\infty} \leq \delta$}
\STATE $\bm{\Lambda}_{i}^{(t + 1)} := \bm{\Lambda}_{i}^{(t)} + \frac{1}{\rho^{(t)}}(\mathbf{U}_{i}^{(t + 1)} - (\mathbf{U}_{i}^{\mathrm{R}})^{(t + 1)})$;
\STATE $\rho^{(t + 1)} := \rho^{(t)}$;
\ELSE
\STATE $\bm{\Lambda}_{i}^{(t + 1)} := \bm{\Lambda}_{i}^{(t)}$, $\rho^{(t + 1)} := c \cdot \rho^{(t)}$;
\ENDIF
\ENDFOR
\STATE $t := t + 1$;
\UNTIL{\textcolor{blue}{a certain stopping criterion is reached}}
\end{algorithmic}
\end{algorithm}

\textcolor{blue}{\textbf{Remark 1.} Observed from simulation results shown in Sec. V, Alg. 1 \& 2 converge and yield nearly identical performance. Nevertheless, when the number of users is small, e.g., $K \leq 3$, Alg. 1 exhibits lower complexity and vice versa. Besides, assuming that Alg. 1 starts from a feasible point and the step size of (projected) GD procedure is sufficiently small, then, it can be proved that the solution iterates generated by Alg. 1 maintain feasible to $(\mathcal{P}1)$ and yields monotonically decreasing objective values of $(\mathcal{P}1)$.}

\section{Performance Analysis and Comparison}

In this section, we analyze the estimation performance of our proposed RIS-TX CE method and compare it with that of the conventional CscdChn training scheme, including both LMMSE and LS estimation. To this end, we employ the DFT matrix as the codebook for pilot sequence, i.e., $\bm{\Phi} = \mathbf{F}$ with $[\mathbf{F}]_{i,j} = e^{-j\frac{2\pi(i - 1)(j - 1)}{M_{\mathrm{R}}}}, \; i,j \in \mathcal{M}$, and uniform power allocation, i.e., $\mathbf{P} = \frac{\sqrt{P_{\mathrm{max}}}}{M_{\mathrm{R}}}\mathbf{I}_{M_{\mathrm{R}}}$. The use of DFT pilot sequence is due to the following considerations:

\begin{itemize}

\item The DFT pilot sequence has been shown to be nearly optimal for the CscdChn training scheme by different works [6]-[8], [10], [11].

\item Although our GD/PDD-algorithm induced pilot sequences yield superior performance over DFT codebook (see Sec. V), utilizing DFT pilot yields closed-form MSE and CRLB expressions, which greatly simplifies analysis and helps obtain more insights.

\end{itemize}

Besides, we also assume all channels follow Rayleigh fading distribution [28], i.e., $\mathbf{g} \sim \mathcal{CN}(\bm{0}, \rho_{\mathrm{G}}\mathbf{I}_{M_{\mathrm{B}}M_{\mathrm{R}}})$ and $\mathbf{h}_{k} \sim \mathcal{CN}(\bm{0}, \rho_{\mathrm{h},k}\mathbf{I}_{M_{\mathrm{R}}}), \; k \in \mathcal{K}$, where $\rho_{\mathrm{G}}$ and $\rho_{\mathrm{h},k}, \; k \in \mathcal{K}$, stand for the large-scale fading coefficients of $\mathbf{G}$ and $\mathbf{h}_{k}$, respectively.

\vspace{-3mm}
\subsection*{A. CscdChn Training Scheme}

To perform analysis and comparison, we first briefly review the conventional CscdChn training procedure, which is conducted in analogy with the classical uplink CE protocol [28]. Specifically, for the system described in Sec. II, all $K$ single-antenna users simultaneously transmit orthogonal pilot sequences to the BS within $T \geq M_{\mathrm{R}}$ consecutive time slots. Here, the direct channels between users and the BS are still not considered since they can be easily obtained in advance [1], [9], [10]. Therefore, in the $t$th time slot, the signal received at the BS is
\begin{equation}
\tilde{\mathbf{Y}}_{\mathrm{B}}^{(t)} = {\sum}_{k=1}^{K}\mathbf{H}_{\mathrm{c},k}\bm{\phi}_{\mathrm{p}}^{(t)}\mathbf{z}_{k}^{\mathrm{H}} + \tilde{\mathbf{N}}_{\mathrm{B}}^{(t)}, \; t \in \mathcal{T},
\end{equation}
where $\mathbf{H}_{\mathrm{c},k}, \; k \in \mathcal{K}$, is the effective CscdChn defined in Sec. II, $\bm{\phi}_{\mathrm{p}}^{(t)}, \; t \in \mathcal{T}$, indicates phase shifts of RIS elements and is set as the $t$th column of the DFT matrix, $\mathbf{z}_{k}, \; k \in \mathcal{K}$, represents pilot sequence transmitted by the $k$th user and $\tilde{\mathbf{N}}_{\mathrm{B}}^{(t)}, \; t \in \mathcal{T}$, denotes the thermal noise at the BS with each element following i.i.d. $\mathcal{CN}(0,\sigma_{\mathrm{B}}^{2})$. We assume that pilot sequences from different users are mutually orthogonal [28], i.e., $\| \mathbf{z}_{k} \|_{2}^{2} = P_{\mathrm{max}}$ and $\mathbf{z}_{k}^{\mathrm{H}}\mathbf{z}_{j} = 0, \; k,j \in \mathcal{K}, \; k \neq j$. To save overhead, we just assume $T = M_{\mathrm{R}}$.

\textbf{Remark 2.} Without channel sparsity assumption, CscdChn scheme has the pilot overhead of at least $M_{\mathrm{R}} + \mathsf{max} \{ K - 1, (K - 1) \lceil M_{\mathrm{R}}/M_{\mathrm{B}} \rceil \}$ [1], [9], which is higher than our proposed RIS-TX scheme (i.e. $M_{\mathrm{R}}$).

To extract each user's channel information, by right multiplying both sides of (52) with $\mathbf{z}_{k}, \; k \in \mathcal{K}$, the BS obtains
\begin{equation}
\tilde{\mathbf{y}}_{\mathrm{B},k}^{(t)} \!=\! \tilde{\mathbf{Y}}_{\mathrm{B}}^{(t)}\mathbf{z}_{k} \!=\! P_{\mathrm{max}}\mathbf{H}_{\mathrm{c},k}\bm{\phi}_{\mathrm{p}}^{(t)} \!+\! \tilde{\mathbf{n}}_{\mathrm{B},k}^{(t)}, k \in \mathcal{K}, t \in \mathcal{T},
\end{equation}
where $\tilde{\mathbf{n}}_{\mathrm{B},k}^{(t)} \triangleq \tilde{\mathbf{N}}_{\mathrm{B}}^{(t)}\mathbf{z}_{k}, \; k \in \mathcal{K}, \; t \in \mathcal{T}$. Via collecting all $T$ $\tilde{\mathbf{y}}_{\mathrm{B},k}^{(t)}$ and stacking them in a tall vector, user-$k$'s data obtained by the BS can be expressed as
\begin{equation}
\tilde{\mathbf{y}}_{\mathrm{B},k} = P_{\mathrm{max}}\mathsf{A}_{\mathrm{G}}(\bm{\Phi}_{\mathrm{p}})\mathbf{h}_{\mathrm{c},k} + \tilde{\mathbf{n}}_{\mathrm{B},k}, \; k \in \mathcal{K},
\end{equation}
where $\bm{\Phi}_{\mathrm{p}} \triangleq [\bm{\phi}_{\mathrm{p}}^{(1)}, \cdots, \bm{\phi}_{\mathrm{p}}^{(T)}]$ is a DFT matrix, $\mathbf{h}_{\mathrm{c},k} \triangleq \mathsf{vec}(\mathbf{H}_{\mathrm{c},k}), \; k \in \mathcal{K}$, and $\tilde{\mathbf{n}}_{\mathrm{B},k} \triangleq \mathsf{vec}\big([\tilde{\mathbf{n}}_{\mathrm{B},k}^{(1)}, \cdots, \tilde{\mathbf{n}}_{\mathrm{B},k}^{(T)}]\big), \; k \in \mathcal{K}$.

Note that $\mathbf{h}_{\mathrm{c},k}, \; k \in \mathcal{K}$, has zero mean and correlation matrix $\mathbf{R}_{\mathrm{H_{c}},k}, \; k \in \mathcal{K}$. Based on (54), adopting the LMMSE estimator, the MSE matrix of $\mathbf{h}_{\mathrm{c},k}, \; k \in \mathcal{K}$, can be shown to be given by [37]
\begin{equation}
\tilde{\mathbf{C}}_{\mathrm{H}_{\mathrm{c},k}} \!\!=\!\! \mathbf{R}_{\mathrm{H_{c}},k} \!\!-\!\! \mathbf{R}_{\mathrm{H_{c}},k}\mathsf{A}_{\mathrm{G}}^{\mathrm{H}}(\!\bm{\Phi}_{\mathrm{p}}\!)\mathbf{J}_{\mathrm{H_{c}},k}\mathsf{A}_{\mathrm{G}}(\!\bm{\Phi}_{\mathrm{p}}\!)\mathbf{R}_{\mathrm{H_{c}},k}, k \!\in\! \mathcal{K},
\end{equation}
where $\mathbf{J}_{\mathrm{H_{c}},k} \triangleq \big(\mathsf{A}_{\mathrm{G}}(\bm{\Phi}_{\mathrm{p}})\mathbf{R}_{\mathrm{H_{c}},k}\mathsf{A}_{\mathrm{G}}^{\mathrm{H}}(\bm{\Phi}_{\mathrm{p}}) + \frac{\sigma_{\mathrm{B}}^{2}}{P_{\mathrm{max}}}\mathbf{I}_{M_{\mathrm{B}}M_{\mathrm{R}}}\big)^{-1}$, $k \in \mathcal{K}$.

\vspace{-3mm}
\subsection*{B. MSE Performance}

Next, we present the estimation performance of the proposed RIS-TX and the classical CscdChn training scheme, including both LMMSE and LS estimation.

\emph{1) LMMSE estimation}

\emph{i) RIS-TX training}

Note the receivers $\mathbf{W}_{\mathrm{G}}$ and $\bm{\mathcal{W}}_{\mathrm{h}}$ are coupled (see (10)) and hence the joint optimal receivers cannot be obtained analytically. To obtain insight, we adopt a suboptimal yet simple receiving scheme---the BS/users conduct LMMSE receiving to minimize CE error of their direct channel to RIS. Specifically, substituting $\mathbf{W}_{\mathrm{G}} = (\mathsf{A}_{\mathrm{G}}(\bm{\Psi})\mathbf{R}_{\mathrm{G}}\mathsf{A}_{\mathrm{G}}^{\mathrm{H}}(\bm{\Psi}) + \bm{\Sigma}_{\mathrm{B}} \otimes \mathbf{I}_{M_{\mathrm{R}}})^{-1}\mathsf{A}_{\mathrm{G}}(\bm{\Psi})\mathbf{R}_{\mathrm{G}}$ and $\mathbf{W}_{\mathrm{h},k} = (\bm{\Psi}^{\mathrm{T}}\mathbf{R}_{\mathrm{h},k}\bm{\Psi}^{*} + \bm{\Sigma}_{\mathrm{U},k})^{-1}\bm{\Psi}^{\mathrm{T}}\mathbf{R}_{\mathrm{h},k}, \; k \in \mathcal{K}$, into (10), the MSE for DFT pilot and Rayleigh fading scenario reduces to (56), as shown on the top of next page.
\begin{figure*}[!t]
\normalsize
\setcounter{MYtempeqncnt}{\value{equation}}
\setcounter{equation}{55}
\begin{equation}
\mathbf{C}_{\mathrm{H}_{\mathrm{c},k}} = \frac{\rho_{\mathrm{G}}^{2}\rho_{\mathrm{h},k}\sigma_{\mathrm{U},k}^{2}M_{\mathrm{R}}P_{\mathrm{max}} + \rho_{\mathrm{h},k}^{2}\rho_{\mathrm{G}}\sigma_{\mathrm{B}}^{2}M_{\mathrm{R}}P_{\mathrm{max}} + \rho_{\mathrm{G}}\rho_{\mathrm{h},k}\sigma_{\mathrm{B}}^{2}\sigma_{\mathrm{U},k}^{2}M_{\mathrm{R}}^{2}}{(\rho_{\mathrm{G}}P_{\mathrm{max}} + \sigma_{\mathrm{B}}^{2}M_{\mathrm{R}})(\rho_{\mathrm{h},k}P_{\mathrm{max}} + \sigma_{\mathrm{U},k}^{2}M_{\mathrm{R}})}\mathbf{I}_{M_{\mathrm{B}}M_{\mathrm{R}}}, \; k \in \mathcal{K}.
\end{equation}
\setcounter{MYtempeqncnt}{\value{equation}}
%\hrulefill
\vspace*{-15pt}
\end{figure*}

\emph{ii) CscdChn training}

Utilizing the DFT pilot and Rayleigh fading hypothesis, the expression in (55) can be simplified into
\begin{equation}
\tilde{\mathbf{C}}_{\mathrm{H}_{\mathrm{c},k}} = \frac{\rho_{\mathrm{G}}\rho_{\mathrm{h},k}\sigma_{\mathrm{B}}^{2}}{\rho_{\mathrm{G}}\rho_{\mathrm{h},k}M_{\mathrm{R}}P_{\mathrm{max}} + \sigma_{\mathrm{B}}^{2}}\mathbf{I}_{M_{\mathrm{B}}M_{\mathrm{R}}}, \; k \in \mathcal{K}.
\end{equation}

\emph{2) LS estimation}

\emph{i) RIS-TX training}

When utilizing LS estimator, i.e., $\mathbf{W}_{\mathrm{G}} = \mathsf{A}_{\mathrm{G}}(\bm{\Psi})(\mathsf{A}_{\mathrm{G}}^{\mathrm{H}}(\bm{\Psi})\mathsf{A}_{\mathrm{G}}(\bm{\Psi}))^{-1}$ and $\mathbf{W}_{\mathrm{h},k} = \bm{\Psi}^{\mathrm{T}}(\bm{\Psi}^{*}\bm{\Psi}^{\mathrm{T}})^{-1}, \; k \in \mathcal{K}$, under the hypothesis of DFT pilot and Rayleigh fading, the equation (10) is simplified to (58), shown on the top of next page.$^{3}$\footnote{$^{3}$Note the channels statistics are used here only for purpose of analysis, which is indeed unavailable when performing LS estimation.}
\begin{figure*}[!t]
\normalsize
\setcounter{MYtempeqncnt}{\value{equation}}
\setcounter{equation}{57}
\begin{equation}
\mathbf{C}_{\mathrm{H}_{\mathrm{c},k}}^{\mathrm{LS}} = \frac{\rho_{\mathrm{h},k}\sigma_{\mathrm{B}}^{2}M_{\mathrm{R}}P_{\mathrm{max}} + \rho_{\mathrm{G}}\sigma_{\mathrm{U},k}^{2}M_{\mathrm{R}}P_{\mathrm{max}} + \sigma_{\mathrm{B}}^{2}\sigma_{\mathrm{U},k}^{2}M_{\mathrm{R}}^{2}}{P_{\mathrm{max}}^{2}}\mathbf{I}_{M_{\mathrm{B}}M_{\mathrm{R}}}, \; k \in \mathcal{K}.
\end{equation}
\setcounter{MYtempeqncnt}{\value{equation}}
\hrulefill
\vspace*{4pt}
\end{figure*}

\emph{ii) CscdChn training}

Applying LS estimation to (54), under the DFT pilots and Rayleigh fading assumptions, the covariance matrix of the CscdChn is given as
\begin{align}
\tilde{\mathbf{C}}_{\mathrm{H}_{\mathrm{c},k}}^{\mathrm{LS}} &= \frac{\sigma_{\mathrm{B}}^{2}}{P_{\mathrm{max}}}((\bm{\Phi}_{\mathrm{p}}^{*}\bm{\Phi}_{\mathrm{p}}^{\mathrm{T}})^{-1} \otimes \mathbf{I}_{M_{\mathrm{B}}})\\
&= \frac{\sigma_{\mathrm{B}}^{2}}{M_{\mathrm{R}}P_{\mathrm{max}}}\mathbf{I}_{M_{\mathrm{B}}M_{\mathrm{R}}}, \; k \in \mathcal{K}.
\end{align}

\subsection*{C. CRLB Analysis}

In this subsection, we present CRLBs for both the RIS-TX and CscdChn training schemes and compare their performance.

\emph{1) CRLB for RIS-TX scheme}

Based on signal model (4), the likelihood function of the random vector $\mathbf{y}_{\mathrm{B}}$ conditioned on $\mathbf{g}$ is
\begin{equation}
\mathsf{f}_{\mathrm{R}}(\mathbf{y}_{\mathrm{B}};\mathbf{g}) = \frac{e^{-\| (\bm{\Sigma}_{\mathrm{B}} \otimes \mathbf{I}_{M_{\mathrm{R}}})^{-\frac{1}{2}}(\mathbf{y}_{\mathrm{B}} - ((\bm{\Phi}\mathbf{P})^{\mathrm{T}} \otimes \mathbf{I}_{M_{\mathrm{B}}})\mathbf{g}) \|_{2}^{2}}}{\pi^{M_{\mathrm{B}}M_{\mathrm{R}}}\mathsf{det}(\bm{\Sigma}_{\mathrm{B}} \otimes \mathbf{I}_{M_{\mathrm{R}}})}.
\end{equation}
The Fisher information matrix (FIM) of $\mathbf{g}$ is given as [37]
\begin{align}
&\mathbf{J}(\mathbf{g}) \triangleq \mathbb{E}\bigg\{ -\frac{\partial^{2}\mathsf{ln}\;\mathsf{f}_{\mathrm{R}}(\mathbf{y}_{\mathrm{B}};\mathbf{g})}{\partial\mathbf{g}^{*}\partial\mathbf{g}} \bigg\} \notag\\
&= ((\bm{\Phi}\mathbf{P})^{*} \otimes \mathbf{I}_{M_{\mathrm{B}}})(\bm{\Sigma}_{\mathrm{B}} \otimes \mathbf{I}_{M_{\mathrm{R}}})^{-1}((\bm{\Phi}\mathbf{P})^{\mathrm{T}} \otimes \mathbf{I}_{M_{\mathrm{B}}}).
\end{align}
Following the similar procedure, the FIM of $\mathbf{h}_{k}, \; k \in \mathcal{K}$, can be expressed as
\begin{equation}
\mathbf{J}(\mathbf{h}_{k}) = (\bm{\Phi}\mathbf{P})^{*}\bm{\Sigma}_{\mathrm{U},k}^{-1}(\bm{\Phi}\mathbf{P})^{\mathrm{T}}, \; k \in \mathcal{K}.
\end{equation}
By exploiting [37, Eq. (3.30)], the CRLB of effective CscdChn's MSE matrix $\mathbf{C}_{\mathrm{H_{c}},k}, \; k \in \mathcal{K}$, is given as follows
\begin{align}
\mathbf{C}_{\mathrm{H_{c}},k} \!\!\succeq\!\! \bigg(\!\frac{\partial\mathbf{h}_{\mathrm{c},k}^{\mathrm{T}}}{\partial\mathbf{g}}\!\bigg)^{\mathrm{H}}\!\!\mathbf{J}^{-\!1}(\!\mathbf{g}\!)\frac{\partial\mathbf{h}_{\mathrm{c},k}^{\mathrm{T}}}{\partial\mathbf{g}} \!\!+\!\! \bigg(\!\frac{\partial\mathbf{h}_{\mathrm{c},k}^{\mathrm{T}}}{\partial\mathbf{h}_{k}}\!\bigg)^{\mathrm{H}}\!\!\mathbf{J}^{-\!1}(\!\mathbf{h}_{k}\!)\frac{\partial\mathbf{h}_{\mathrm{c},k}^{\mathrm{T}}}{\partial\mathbf{h}_{k}},
\end{align}
where
\begin{align}
&\frac{\partial\mathbf{h}_{\mathrm{c},k}^{\mathrm{T}}}{\partial\mathbf{g}} = \mathsf{Diag}(\mathbf{h}_{k}) \otimes \mathbf{I}_{M_{\mathrm{B}}}, \; k \in \mathcal{K}, \notag\\
&\frac{\partial\mathbf{h}_{\mathrm{c},k}^{\mathrm{T}}}{\partial\mathbf{h}_{k}} = \mathsf{blkdiag}([\mathbf{G}]_{:,1}^{\mathrm{T}}, \cdots, [\mathbf{G}]_{:,M_{\mathrm{R}}}^{\mathrm{T}}), \; k \in \mathcal{K}.
\end{align}

Utilizing DFT pilot and assuming Rayleigh fading channels, the NMSE of the RIS-TX scheme is lower-bounded by
\begin{equation}
\mathsf{NMSE}_{k} \!\!\triangleq\!\! \frac{\mathsf{Tr} \{\! \mathbf{C}_{\mathrm{H}_{\mathrm{c},k}} \!\}}{\mathsf{Tr} \{\! \mathbf{R}_{\mathrm{H}_{\mathrm{c},k}} \!\}} \!\!\geq\!\! \frac{\sigma_{\mathrm{B}}^{2}M_{\mathrm{B}}\| \mathbf{h}_{k} \|_{2}^{2} \!+\! \sigma_{\mathrm{U},k}^{2}\| \mathbf{G} \|_{\mathrm{F}}^{2}}{\rho_{\mathrm{G}}\rho_{\mathrm{h},k}M_{\mathrm{B}}P_{\mathrm{max}}}, k \!\in\! \mathcal{K},
\end{equation}
where $\mathsf{Tr} \{ \mathbf{R}_{\mathrm{H}_{\mathrm{c},k}} \} = \rho_{\mathrm{G}}\rho_{\mathrm{h},k}M_{\mathrm{B}}M_{\mathrm{R}}, \; k \in \mathcal{K}$. Note the above CRLB is dependent on instant realization of $\mathbf{h}_{k}, \; k \in \mathcal{K}$, and $\mathbf{G}$. To eliminate the randomness deriving from channels, taking expectations w.r.t. channels, we obtain
\begin{equation}
\mathsf{NMSE}_{k} \geq \frac{\rho_{\mathrm{h},k}\sigma_{\mathrm{B}}^{2}M_{\mathrm{R}} + \rho_{\mathrm{G}}\sigma_{\mathrm{U},k}^{2}M_{\mathrm{R}}}{\rho_{\mathrm{G}}\rho_{\mathrm{h},k}P_{\mathrm{max}}}, \; k \in \mathcal{K}.
\end{equation}

\emph{2) CRLB for CscdChn scheme}

The likelihood function of the receiving signal in (54) given $\mathbf{h}_{\mathrm{c},k}, \; k \in \mathcal{K}$, is
\begin{equation}
\mathsf{f}_{\mathrm{C}}(\tilde{\mathbf{y}}_{\mathrm{B},k};\mathbf{h}_{\mathrm{c},k}) = \frac{e^{-\frac{\| \tilde{\mathbf{y}}_{\mathrm{B},k} - P_{\mathrm{max}}(\bm{\Phi}_{\mathrm{p}}^{\mathrm{T}} \otimes \mathbf{I}_{M_{\mathrm{B}}})\mathbf{h}_{\mathrm{c},k} \|_{2}^{2}}{\sigma_{\mathrm{B}}^{2}P_{\mathrm{max}}}}}{(\pi\sigma_{\mathrm{B}}^{2}P_{\mathrm{max}})^{M_{\mathrm{B}}M_{\mathrm{R}}}}, \; k \in \mathcal{K}.
\end{equation}
The FIM of $\mathbf{h}_{\mathrm{c},k}, \; k \in \mathcal{K}$, can be readily obtained as
\begin{equation}
\tilde{\mathbf{J}}(\mathbf{h}_{\mathrm{c},k}) = \frac{P_{\mathrm{max}}}{\sigma_{\mathrm{B}}^{2}}((\bm{\Phi}_{\mathrm{p}}^{*}\bm{\Phi}_{\mathrm{p}}^{\mathrm{T}}) \otimes \mathbf{I}_{M_{\mathrm{B}}}), \; k \in \mathcal{K}.
\end{equation}

Under the DFT pilot and Rayleigh fading channel assumptions, the CRLB of the CscdChn scheme's NMSE is
\begin{equation}
\mathsf{NMSE}_{k} \geq \frac{\mathsf{Tr} \{ \tilde{\mathbf{J}}^{-1}(\mathbf{h}_{\mathrm{c},k}) \}}{\mathsf{Tr} \{ \mathbf{R}_{\mathrm{H_{c}},k} \}} = \frac{\sigma_{\mathrm{B}}^{2}}{\rho_{\mathrm{G}}\rho_{\mathrm{h},k}M_{\mathrm{R}}P_{\mathrm{max}}}, \; k \in \mathcal{K}.
\end{equation}

\emph{3) RIS-TX v.s. CscdChn --- The Watershed $M_{\mathrm{R}}$}

Comparing (67) and (70), we notice one important fact is that $M_{\mathrm{R}}$ exerts opposite impact on the two competing CE schemes. The growth of $M_{\mathrm{R}}$ increases the RIS-TX scheme's CRLB while lowering that of the CscdChn method, which implies that there exists a threshold $M_{\mathrm{R}}^{\mathrm{Th}}$ such that the RIS-TX scheme outperforms CscdChn when $M_{\mathrm{R}} < M_{\mathrm{R}}^{\mathrm{Th}}$ and vice versa. In the following, we investigate the impact of $P_{\mathrm{max}}$ and pathloss fading on the cross-point $M_{\mathrm{R}}^{\mathrm{Th}}$.

\begin{figure} [!t]
\centering
\includegraphics[scale=0.60]{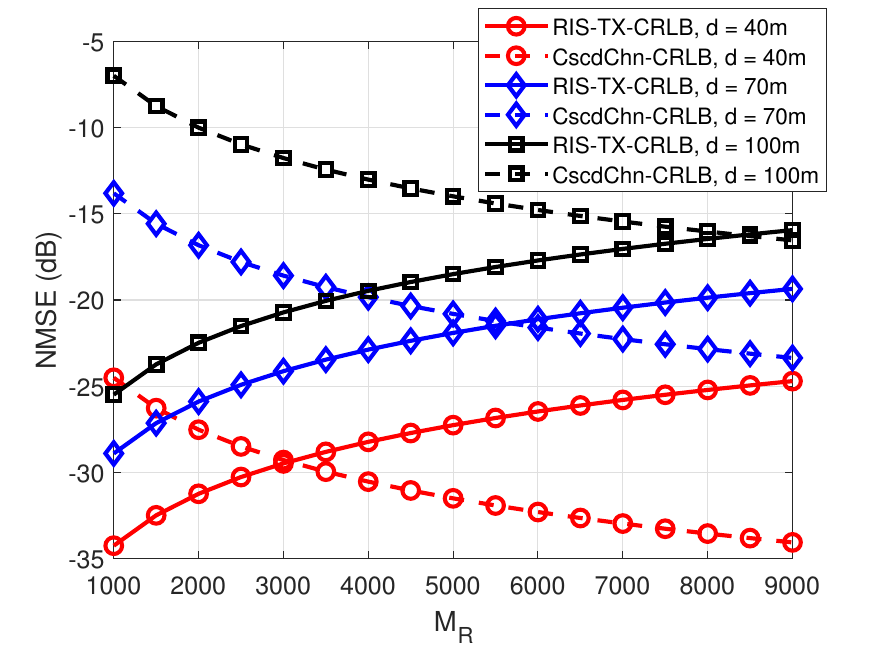}
\caption*{Fig. 2. The impact of pathloss fading on $M_{\mathrm{R}}^{\mathrm{Th}}$ ($P_{\mathrm{max}} = 50\mathrm{dBm}$, $\{ \sigma_{\mathrm{U},k}^{2} \}_{k=1}^{K} = \sigma_{\mathrm{B}}^{2} = -90\mathrm{dBm}$).}
\end{figure}

\emph{i) Impact of $P_{\mathrm{max}}$ on $M_{\mathrm{R}}^{\mathrm{Th}}$}

With other system settings fixed, $M_{\mathrm{R}}^{\mathrm{Th}}$ will not change when $P_{\mathrm{max}}$ varies. In fact, via dividing the rightmost side of (67) by that of (70), the ratio $\frac{(\sigma_{\mathrm{B}}^{2}M_{\mathrm{B}}\| \mathbf{h}_{k} \|_{2}^{2} + \sigma_{\mathrm{U},k}^{2}\| \mathbf{G} \|_{\mathrm{F}}^{2})M_{\mathrm{R}}}{\sigma_{\mathrm{B}}^{2}M_{\mathrm{B}}}$ is independent of $P_{\mathrm{max}}$.

\emph{ii) Impact of pathloss fading on $M_{\mathrm{R}}^{\mathrm{Th}}$}

To investigate the impact of pathloss on $M_{\mathrm{R}}^{\mathrm{Th}}$, we adopt the classical 3GPP Urban Micro channel model [38] in the following comparison. Specifically, according to [38], the channel fading gain for a typical indoor environment is given as
\begin{equation}
\mathsf{f}(d) = G_{\mathrm{t}} + G_{\mathrm{r}} - 37.5 - 22\log_{10}(d/1\mathrm{m}) \; [\mathrm{dB}],
\end{equation}
where $G_{\mathrm{t}}$ and $G_{\mathrm{r}}$ indicate the antenna gains (in dBi) at the transmitter and the receiver, respectively, and $d$ is the distance (in meters) between the TX and RX terminals (here $d$ is the distance between the RIS and BS/users). Besides, we denote $\{ \sigma_{\mathrm{U},k}^{2} \}_{k=1}^{K} = \sigma_{\mathrm{B}}^{2} = \sigma^{2}$ and omit the subscript $k$ in (67) and (70) for simplicity. In Fig. 2, under various pathloss settings (i.e. $d$), we present CRLBs of two CE schemes and the associated cross-point $M_{\mathrm{R}}^{\mathrm{Th}}$. The results in Fig. 2 indicate that the severer pathloss is, the larger $M_{\mathrm{R}}^{\mathrm{Th}}$ is. The underlying reason lies in that CscdChn scheme experiences “double”-fading while RIS-TX has only “single”-fading, which makes CscdChn scheme influenced by pathloss fading much more severely.

Especially, as shown in Fig. 2, for typical indoor scenarios, the CscdChn scheme outperforms its RIS-TX counterpart when RIS has large size (i.e. $M_{\mathrm{R}} \geq 3000$), which implies that our RIS-TX scheme can be beneficial for RIS device with small/moderate size.

\textbf{Remark 3.} Note that the above discussion on $M_{\mathrm{R}}^{\mathrm{Th}}$ is based on CRLB, which is only achievable when $P_{\mathrm{max}}$ is extremely large. For finite $P_{\mathrm{max}}$, the cross-point $M_{\mathrm{R}}^{\mathrm{Th}}$ can be unrealistically large, e.g., $\geq$ 10000 (see Sec. IV. E).

\subsection*{D. Scaling Laws}

In the following, we examine the scaling laws of the RIS-TX and CscdChn schemes when $P_{\mathrm{max}}$ and $M_{\mathrm{R}}$ are in limit regime. Here, we can focus on the CE performance of one specific user $k$. The normalized MSE (NMSE) is defined in (66), where $\mathsf{Tr} \{ \mathbf{R}_{\mathrm{H}_{\mathrm{c},k}} \} = \rho_{\mathrm{G}}\rho_{\mathrm{h},k}M_{\mathrm{B}}M_{\mathrm{R}}, \; k \in \mathcal{K}$, with Rayleigh fading assumption. For simplicity, we denote $\{ \rho_{\mathrm{h},k} \}_{k=1}^{K} = \rho_{\mathrm{h}}$ and $\sigma_{\mathrm{U},k}^{2} = \sigma_{\mathrm{B}}^{2} = \sigma^{2}$.

\begin{figure} [!t]
\centering
\includegraphics[scale=0.60]{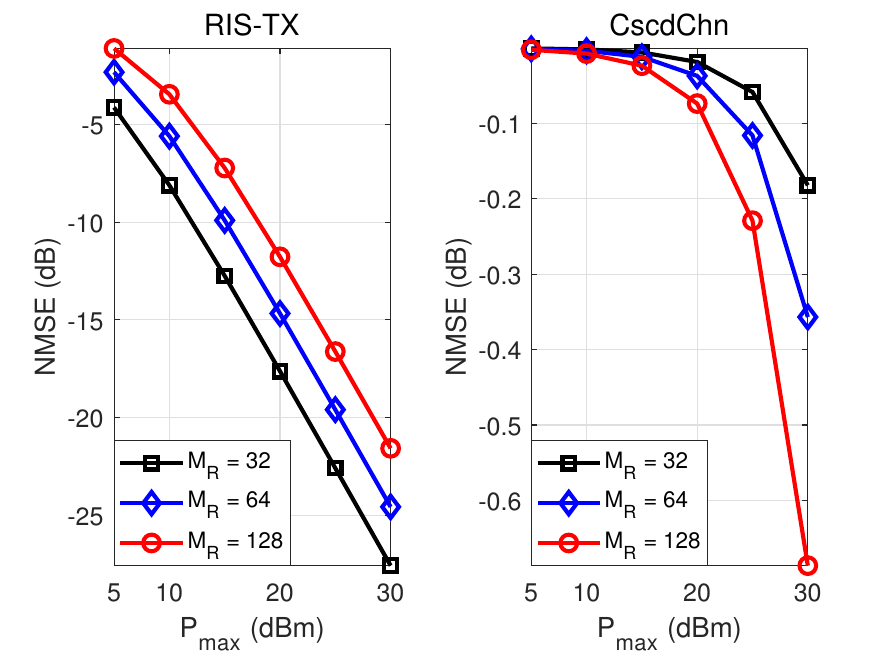}
\caption*{Fig. 3. RIS-TX training v.s. CscdChn training: Impact of $P_{\mathrm{max}}$ (LMMSE estimation).}
\end{figure}

\emph{1) Scaling law of $P_{\mathrm{max}}$}

Define $\mathsf{SNR} = P_{\mathrm{max}}/\sigma^{2}$. Then, when $\mathsf{SNR}$ is large, it can be shown that the NMSE of the RIS-TX training scheme, including both LMMSE and LS estimation, can be approximated by
\begin{equation}
\mathsf{NMSE} \approx \frac{\rho_{\mathrm{G}}M_{\mathrm{R}} + \rho_{\mathrm{h}}M_{\mathrm{R}}}{\rho_{\mathrm{G}}\rho_{\mathrm{h}}}\mathcal{O}(\mathsf{SNR}^{-1}).
\end{equation}
At the same time, the NMSE of the CscdChn scheme using either LMMSE or LS converges to
\begin{equation}
\mathsf{NMSE} \approx \frac{1}{\rho_{\mathrm{G}}\rho_{\mathrm{h}}M_{\mathrm{R}}}\mathcal{O}(\mathsf{SNR}^{-1}).
\end{equation}

As seen above, when $P_{\mathrm{max}}$ is large, the NMSE of the two competing training schemes both converge to zero, but with different rates. For realistic setting, where $M_{\mathrm{R}}$ is finite (several tens or hundreds) and the distances of BS-RIS and RIS-user are over ten meters, CscdChn generally converges faster.

\begin{figure} [!t]
\centering
\includegraphics[scale=0.60]{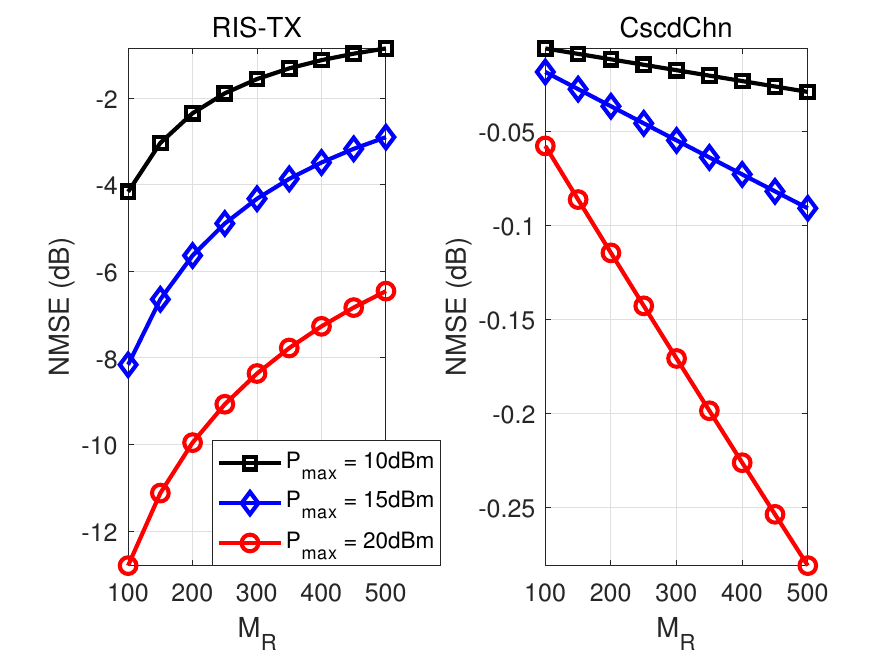}
\caption*{Fig. 4. RIS-TX training v.s. CscdChn training: Impact of $M_{\mathrm{R}}$ (LMMSE estimation).}
\end{figure}

\begin{figure} [!t]
\centering
\includegraphics[scale=0.60]{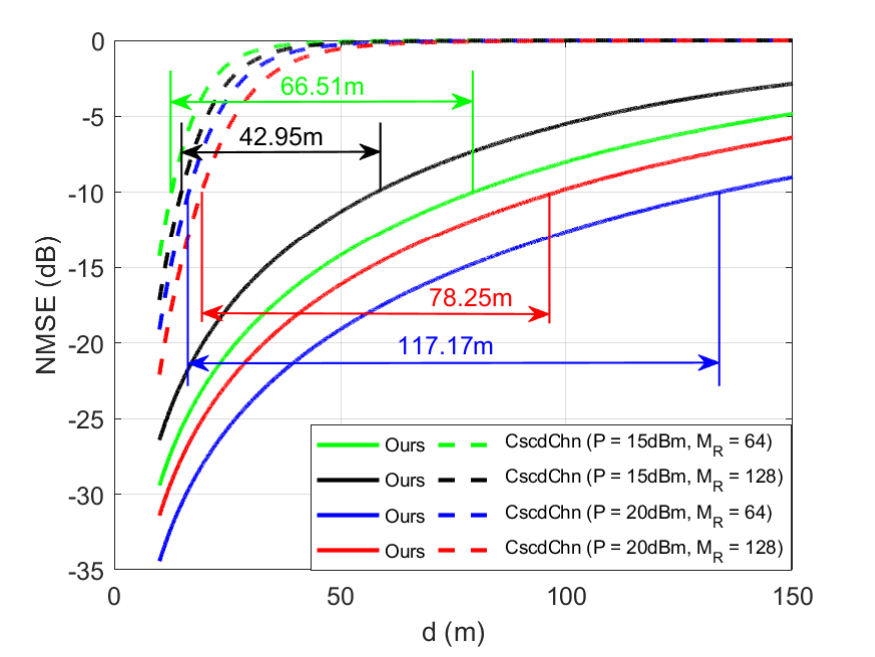}
\caption*{Fig. 5. RIS-TX training v.s. CscdChn training: Coverage (LMMSE estimation).}
\end{figure}

\emph{2) Scaling law of $M_{\mathrm{R}}$}

With infinite $M_{\mathrm{R}}$, it can be observed that the NMSE of the RIS-TX training method exploiting LMMSE estimation converges to $\frac{\rho_{\mathrm{G}}\rho_{\mathrm{h}}\sigma^{4}}{\sigma^{4}} \times \frac{M_{\mathrm{B}}M_{\mathrm{R}}}{\rho_{\mathrm{G}}\rho_{\mathrm{h}}M_{\mathrm{B}}M_{\mathrm{R}}} = 1$, while that utilizing LS estimator approaches to
\begin{equation}
\mathsf{NMSE} \approx \frac{1}{\rho_{\mathrm{G}}\rho_{\mathrm{h}}\mathsf{SNR}^{2}}\mathcal{O}(M_{\mathrm{R}}^{2}) \xrightarrow{M_{\mathrm{R}} \rightarrow \infty} \infty.
\end{equation}
Meanwhile, the NMSE of the CscdChn scheme leveraging both LMMSE and LS estimation converges to
\begin{equation}
\mathsf{NMSE} \approx \frac{1}{\rho_{\mathrm{G}}\rho_{\mathrm{h}}\mathsf{SNR}}\mathcal{O}(M_{\mathrm{R}}^{-1}) \xrightarrow{M_{\mathrm{R}} \rightarrow \infty} 0.
\end{equation}

As demonstrated above, with the RIS size extremely large, the NMSE of CscdChn training scheme converges to zero utilizing both LMMSE and LS estimation, which always outperforms the RIS-TX estimation method because the latter converges to a constant (LMMSE estimator) or tends to be infinite (LS estimator). The reason lies in that, when $M_{\mathrm{R}}$ is large, more element-wise CscdChns will be combined to form the effective channel in CscdChn estimation scheme, which indeed enlarges the effective channel's magnitude and therefore improves the CE performance. In contrast, for RIS-TX training scheme, when $M_{\mathrm{R}}$ grows, the average power allocated to each element-wise channel decreases, which deteriorates the CE precision. Therefore, $M_{\mathrm{R}}$ exerts opposite influence onto the two training schemes.

The above asymptotic analysis appears to suggest that the CscdChn training can be more beneficial than the RIS-TX scheme. However, in practice, realistic settings of $P_{\mathrm{max}}$ and $M_{\mathrm{R}}$ are in finite regime, as elaborated in the next subsection.

\vspace{-3.5mm}
\subsection*{E. CE Performance for Realistic Scenarios}

\vspace{-0.5mm}
We continue to assess different training schemes' performance in realistic settings. To this end, we still adopt the classical 3GPP Urban Micro channel model [38] in the following comparison. Based on (71), we consider a typical indoor setting that both BS-RIS and RIS-user distances are set as $d = 80\mathrm{m}$ and $\sigma_{\mathrm{U},k}^{2} = \sigma_{\mathrm{B}}^{2} = -90\mathrm{dBm}$ in the following assessment.

\emph{1) Comparison of $P_{\mathrm{max}}$}

Based on the above setting, we visually compare the two training schemes using LMMSE by plotting their NMSE curves w.r.t. $P_{\mathrm{max}}$ in Fig. 3, whose left and right half illustrates the RIS-TX and CscdChn scheme, respectively. As shown by Fig. 3, RIS-TX scheme significantly outperforms the CscdChn counterpart. For instance, by setting $M_{\mathrm{R}} = 64$, to achieve NMSE of $-10\mathrm{dB}$, the TX power of RIS-TX is $0.032\mathrm{W}$ while that for the CscdChn counterpart reaches up to $105.1\mathrm{W}$! When $P_{\mathrm{max}}$ is $20\mathrm{dBm}$, RIS-TX is $14.64\mathrm{dB}$ superior to CscdChn in NMSE!

\emph{2) Comparison of $M_{\mathrm{R}}$}

We present NMSE w.r.t. $M_{\mathrm{R}}$ of RIS-TX and CscdChn schemes in the left and right part of Fig. 4, respectively, where LMMSE estimation is used. As seen, increasing $M_{\mathrm{R}}$ imposes opposite impacts onto the two training schemes. Although growing $M_{\mathrm{R}}$ can benefit the CscdChn scheme, its $M_{\mathrm{R}}$ has to be prohibitively large to beat the RIS-TX opponent. For instance, with $P_{\mathrm{max}} = 20\mathrm{dBm}$, to reach $-10\mathrm{dB}$ NMSE level, $M_{\mathrm{R}}$ has to exceed \underline{$6.73 \times 10^{4}$}, which is unrealistic.

\emph{3) Comparison of coverage}

We also examine the coverage of the two training schemes in the above realistic setting. The two schemes' NMSE curves w.r.t. the distance $d$ are presented in Fig. 5, where LMMSE estimation is used. As seen, by fixing $P_{\mathrm{max}} = 15\mathrm{dBm}$ and $M_{\mathrm{R}} = 64$, to reach $-10\mathrm{dB}$ NMSE level, the coverage of CscdChn scheme is $12.67\mathrm{m}$, while that of the RIS-TX can extend up to $79.18\mathrm{m}$.

\vspace{-2mm}
\section{Simulation Results}

\begin{figure} [!t]
\centering
\includegraphics[scale=0.40]{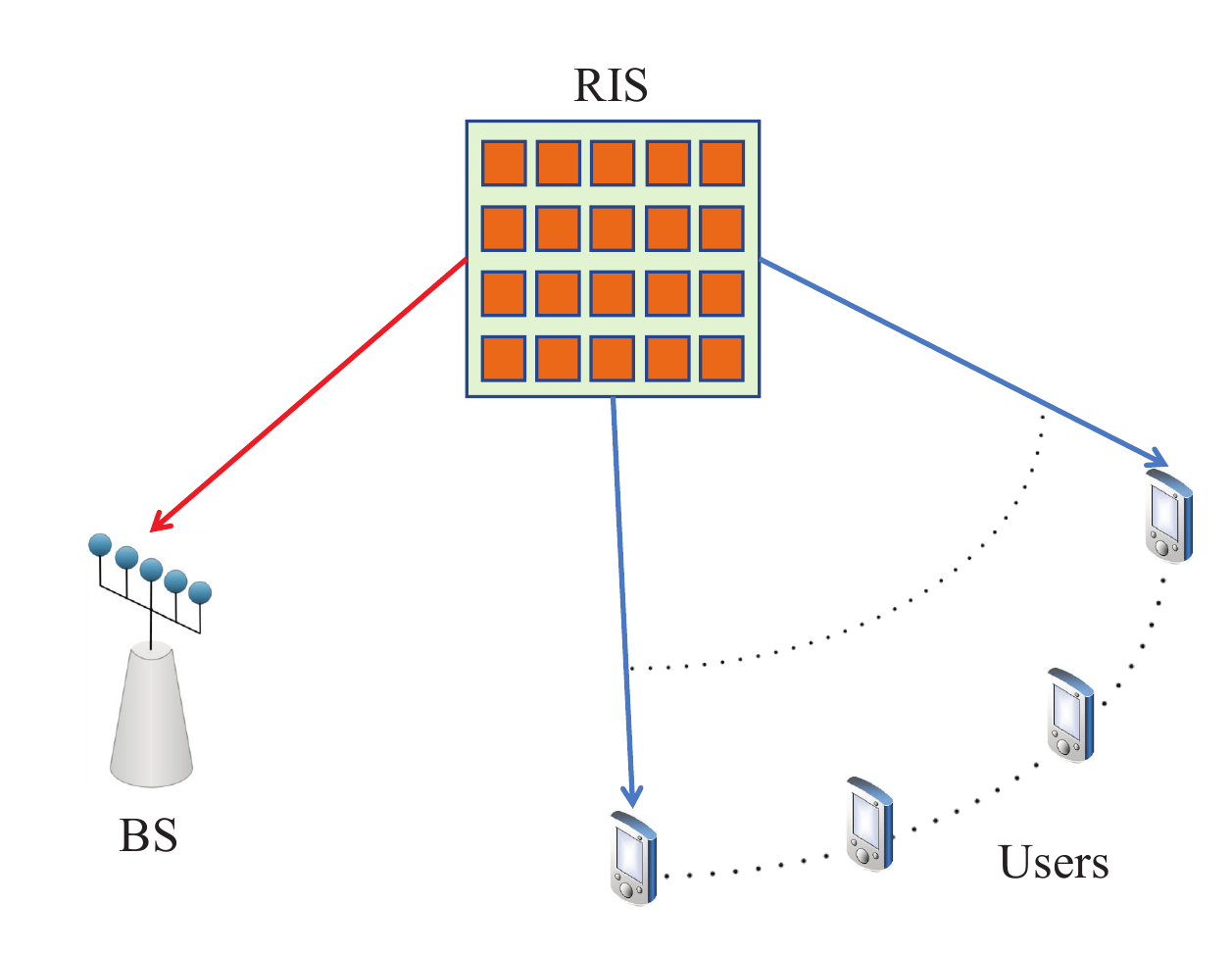}
\caption*{Fig. 6. Simulation scenario.}
\end{figure}

\begin{figure} [!t]
\centering
\includegraphics[scale=0.60]{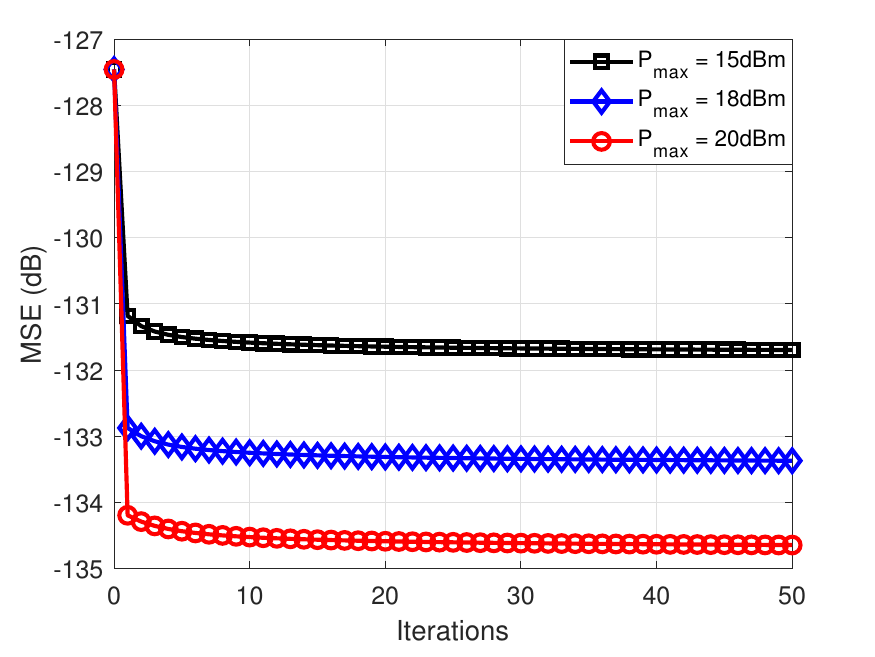}
\caption*{Fig. 7. Convergence of GD algorithm (Alg. 1).}
\end{figure}

In this section, extensive numerical results will be presented to verify the effectiveness of our algorithms and the benefit of the proposed RIS-TX training scheme. As illustrated in Fig. 6, we consider a system which includes a BS with $M_{\mathrm{B}} = 8$ antennas, $K = 4$ users and a RIS equipped with one TX RF-chain. The covariance matrices are given by $\mathbf{R}_{\mathrm{G}} = \rho_{\mathrm{G}}(\mathbf{R}_{\mathrm{GR}} \otimes \mathbf{R}_{\mathrm{GB}})$ and $\mathbf{R}_{\mathrm{h},k} = \rho_{\mathrm{h},k}\mathbf{R}_{\mathrm{hR},k}, \; k \in \mathcal{K}$, where $\mathbf{R}_{\mathrm{GR}}$, $\mathbf{R}_{\mathrm{GB}}$ and $\mathbf{R}_{\mathrm{hR},k}, \; k \in \mathcal{K}$, denote the spatial correlation matrices at the RIS for $\mathbf{G}$, at the BS for $\mathbf{G}$ and at the RIS for $\mathbf{h}_{k}, \; k \in \mathcal{K}$, respectively. Hence, the covariance matrix of $\mathbf{h}_{\mathrm{c},k}, \; k \in \mathcal{K}$, can be calculated as $\mathbf{R}_{\mathrm{c},k} = \rho_{\mathrm{G}}\rho_{\mathrm{h},k}((\mathbf{R}_{\mathrm{hR},k} \odot \mathbf{R}_{\mathrm{GR}}) \otimes \mathbf{R}_{\mathrm{GB}}), \; k \in \mathcal{K}$ [1]. Unless otherwise specified, the distance-dependent pathloss is established as (71), $d_{\mathrm{B}} = \{ d_{\mathrm{U},k} \}_{k=1}^{K} = d = 100\mathrm{m}$, $\bm{\Sigma}_{\mathrm{B}} = \sigma_{\mathrm{B}}^2\mathbf{I}_{M_{\mathrm{B}}}$, $\sigma_{\mathrm{B}}^{2} = -90\mathrm{dBm}$ and $\sigma_{\mathrm{U},k}^{2} = -80\mathrm{dBm}, \; k \in \mathcal{K}$.

\begin{figure} [!t]
\centering
\includegraphics[scale=0.60]{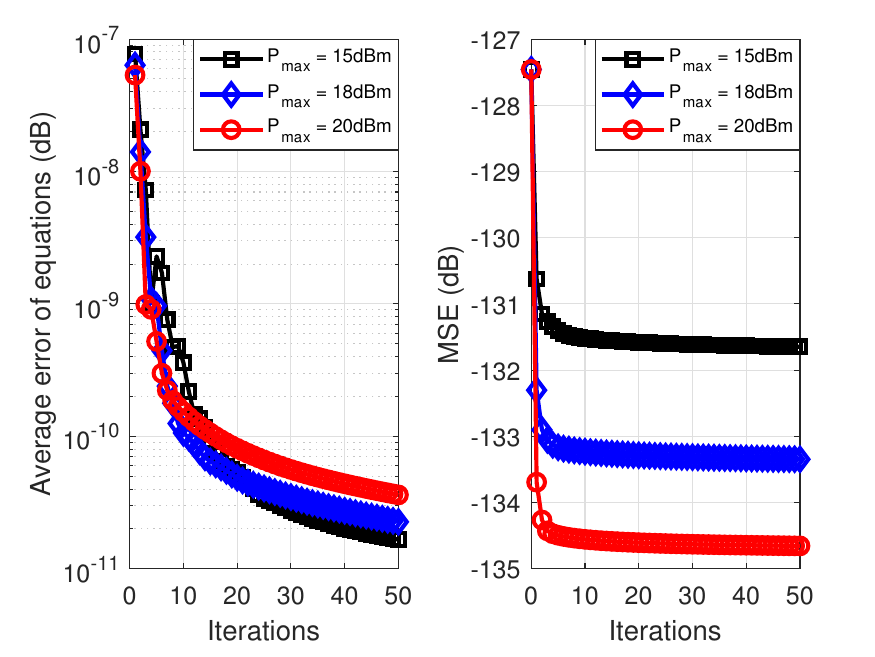}
\caption*{Fig. 8. Convergence of PDD algorithm (Alg. 2).}
\end{figure}

\begin{figure} [!t]
\centering
\includegraphics[scale=0.60]{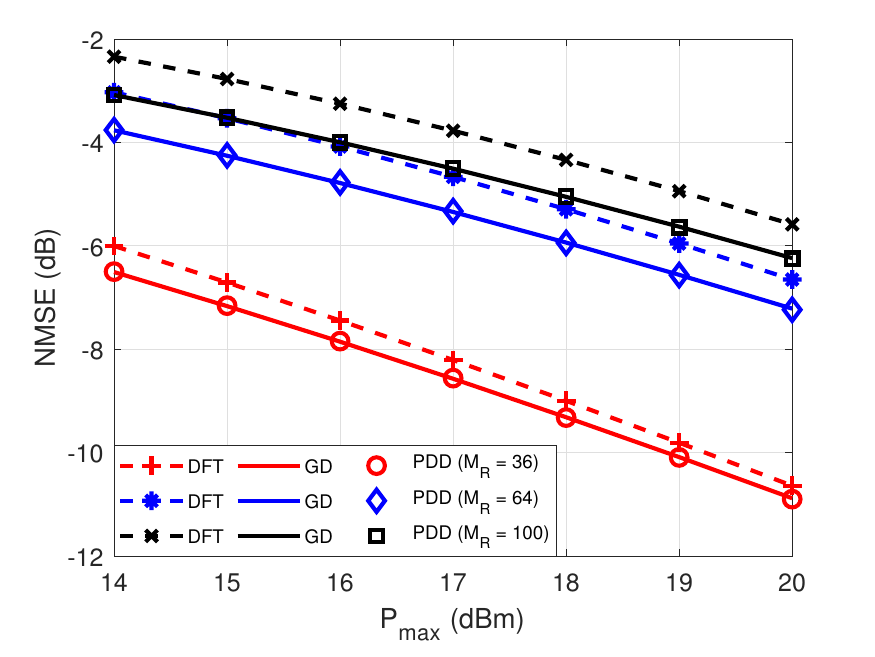}
\caption*{Fig. 9. The impact of $P_{\mathrm{max}}$ on CE performance.}
\end{figure}

Fig. 7 illustrates the convergence of Alg. 1. It can be observed that, the GD algorithm generally converges within two or three tens of iterations.

Fig. 8 verifies the convergence behaviours of the proposed PDD algorithm. The left part of Fig. 8 illustrates the violation of the equalities in (24)-(27) and the right part illustrates objective iterates. The PDD algorithm generally converges very fast.

In Fig. 9, we investigate the impact of the transmit power $P_{\mathrm{max}}$ on the MSE performance. As suggested by the figure, the MSE decreases when the transmit power grows. Interestingly, the pilots numerical designed by our proposed GD and PDD solutions yield nearly identical performance. Importantly, the results in Fig. 9 reflect that the pilot sequence yielded by our proposed GD/PDD algorithms outperforms the classical DFT codebook for RIS-TX training scheme. Note that the DFT sequence is nearly optimal for the classical cascaded CE scheme, as demonstrated by different literature [6]-[8], [10], [11].

Fig. 10 plots the impact of the number of reflecting elements on the CE performance. As shown in Fig. 10, given a specific $P_{\mathrm{max}}$, the CE performance degrades when the number of RIS elements $M_{\mathrm{R}}$ increases, whose reason has been presented in Sec. IV. D. Besides, as previously discussed, the pilot sequences yielded by our GD/PDD algorithms have superior performance than the DFT counterpart.

\begin{figure} [!t]
\centering
\includegraphics[scale=0.60]{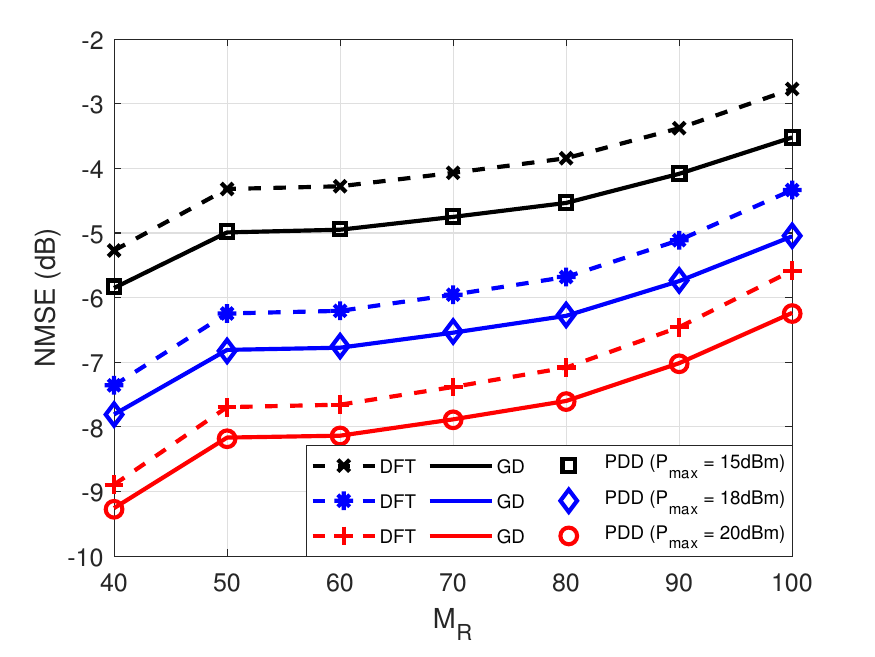}
\caption*{Fig. 10. The impact of $M_{\mathrm{R}}$ on CE performance.}
\end{figure}

\begin{figure} [!t]
\centering
\includegraphics[scale=0.60]{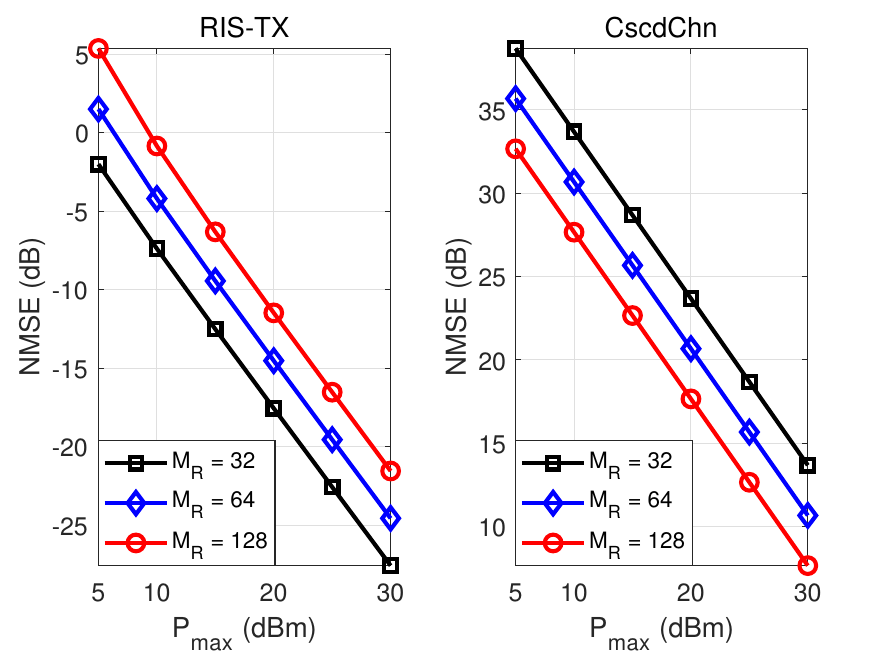}
\caption*{Fig. 11. RIS-TX training v.s. CscdChn training: Impact of $P_{\mathrm{max}}$ (LS estimation, $d = 80\mathrm{m}$, $\sigma_{\mathrm{U},k}^{2} = \sigma_{\mathrm{B}}^{2} = -90\mathrm{dBm}$).}
\end{figure}

\begin{figure} [!t]
\centering
\includegraphics[scale=0.60]{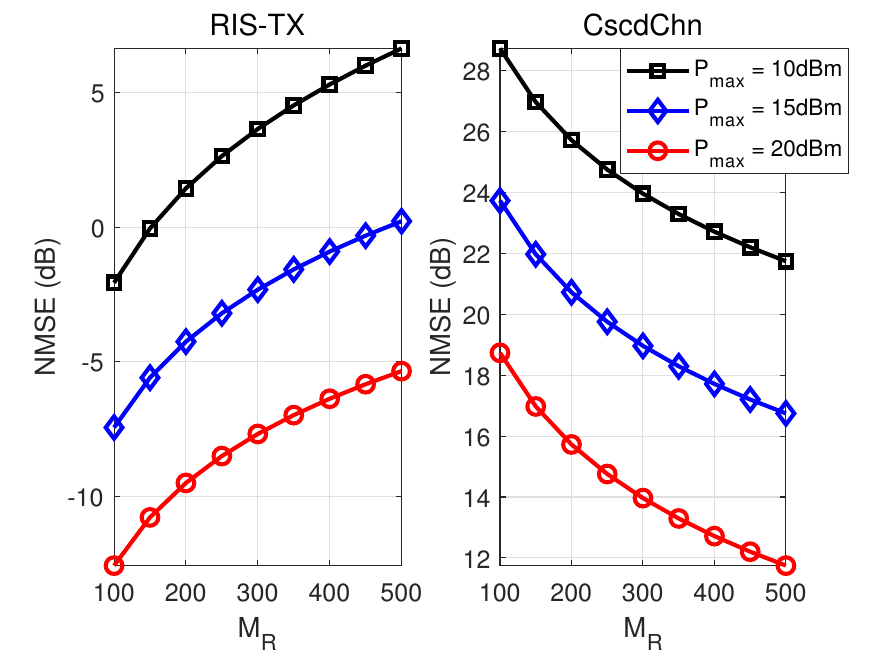}
\caption*{Fig. 12. RIS-TX training v.s. CscdChn training: Impact of $M_{\mathrm{R}}$ (LS estimation, $d = 80\mathrm{m}$, $\sigma_{\mathrm{U},k}^{2} = \sigma_{\mathrm{B}}^{2} = -90\mathrm{dBm}$).}
\end{figure}

\begin{figure} [!t]
\centering
\includegraphics[scale=0.60]{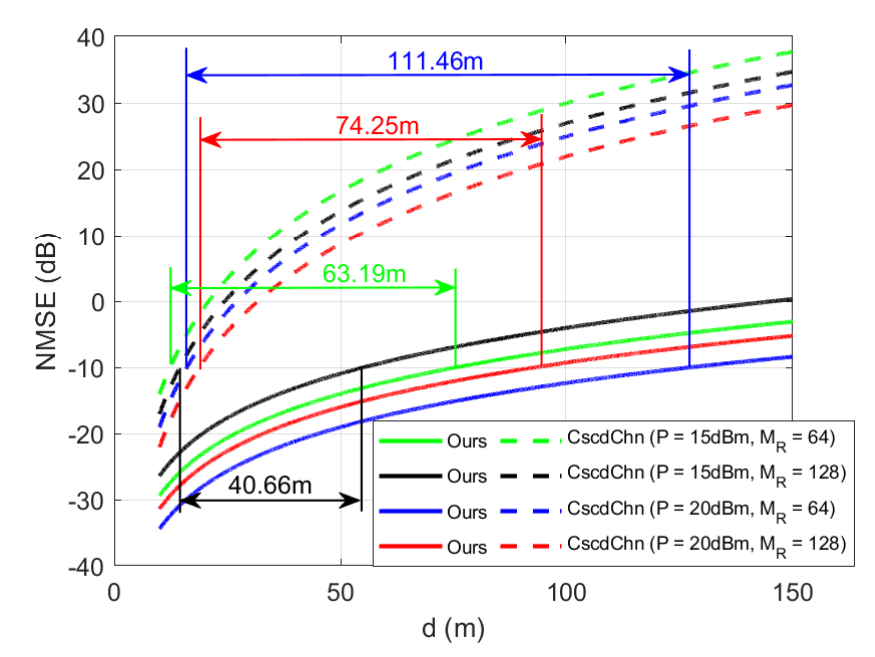}
\caption*{Fig. 13. RIS-TX training v.s. CscdChn training: Coverage (LS estimation, $\sigma_{\mathrm{U},k}^{2} = \sigma_{\mathrm{B}}^{2} = -90\mathrm{dBm}$).}
\end{figure}

In Fig. 11, we compare the MSE performance of our proposed RIS-TX CE scheme and the standard CscdChn training scheme [8], [9] based on LS estimation. Specifically, the left half presents CE performance associated with the RIS-TX training scheme while the right half plots that of the standard CscdChn training. As suggested by the results in Fig. 11, the proposed RIS-TX scheme significantly outperforms the CscdChn training. This is because the severe double fading effect severely attenuates the power of CscdChn. At the same time, it is interesting to note that increasing $M_{\mathrm{R}}$ tends to improve the CE performance for CscdChn training scheme, as opposed to the impact observed in RIS-TX scheme (recall Fig. 10). Detailed explanation of this phenomenon has been given previously. At the same time, LMMSE estimator dramatically surpasses the LS counterpart, since the latter one does not suppress the noise, which is indeed much larger than the pilot signal (recall Fig. 3).

In Fig. 12, the impact of $M_{\mathrm{R}}$ on CE precision is illustrated for LS. The left half exhibits the NMSE performance of RIS-TX training scheme and the right half presents those for the CscdChn training. As shown by this figure, for the test case that corresponds to typical indoor setting, the RIS-TX scheme is obviously superior over the CscdChn training. Besides, as previously explained, increment of $M_{\mathrm{R}}$ exerts opposite influence on NMSE of the two channel training schemes.

In Fig. 13, we investigate the coverage of the proposed RIS-TX and CscdChn training schemes using LS estimator. Based on the pathloss modeling specified in [38] (see (71)), we plot the relation between NMSE and the distance $d$ in Fig. 13. To reach a reasonable NMSE level of $-10\mathrm{dB}$, we indicates coverage difference between the two channel training schemes in the plots. As can be seen, the proposed RIS-TX scheme can enlarge the coverage by several tens or even a hundred meters in an typical indoor scenario.

\section{Conclusion}

In this paper, we propose a novel RIS-TX channel training scheme, where the RIS is equipped with one TX RF-chain and is able to transmit pilot signals during the training period. The new training scheme's pilot sequence design problem turns out to be a very challenging quartic optimization problem. We have developed two efficient solutions to design pilot sequences. Both of these algorithms yield pilots with superior estimation performance over the DFT sequence, which is widely believed to be nearly optimal in the literature. Both theoretic analysis and numerical experiment results show that our proposed RIS-TX training scheme can significantly outperform the classical CscdChn training method in realistic scenarios.

\appendix

\subsection*{A. Derivation of (10)}

According to the definition of $\mathbf{C}_{\mathrm{H}_{\mathrm{c}},k}, \; k \in \mathcal{K}$, its value can be calculated as
\begin{align}
&\mathbf{C}_{\mathrm{H}_{\mathrm{c}},k} = \mathbb{E} \{ \mathsf{vec}(\mathbf{G}\mathsf{Diag}(\mathbf{h}_{k}))\mathsf{vec}^{\mathrm{H}}(\mathbf{G}\mathsf{Diag}(\mathbf{h}_{k})) \} \notag\\
&+ \mathbb{E} \{ \mathsf{vec}(\hat{\mathbf{G}}\mathsf{Diag}(\hat{\mathbf{h}}_{k}))\mathsf{vec}^{\mathrm{H}}(\hat{\mathbf{G}}\mathsf{Diag}(\hat{\mathbf{h}}_{k})) \} \notag\\
&- \mathbb{E} \{ \mathsf{vec}(\mathbf{G}\mathsf{Diag}(\mathbf{h}_{k}))\mathsf{vec}^{\mathrm{H}}(\hat{\mathbf{G}}\mathsf{Diag}(\hat{\mathbf{h}}_{k})) \} \notag\\
&- \mathbb{E} \{ \mathsf{vec}(\hat{\mathbf{G}}\mathsf{Diag}(\hat{\mathbf{h}}_{k}))\mathsf{vec}^{\mathrm{H}}(\mathbf{G}\mathsf{Diag}(\mathbf{h}_{k})) \}, \; k \in \mathcal{K}.
\end{align}

We firstly evaluate the first two expectations. To this end, we study the following expectation
\begin{equation}
\mathbf{R} = \mathbb{E} \{ \mathsf{vec}(\mathbf{X}\mathsf{Diag}(\mathbf{y}))\mathsf{vec}^{\mathrm{H}}(\mathbf{X}\mathsf{Diag}(\mathbf{y})) \},
\end{equation}
where $\mathbf{X} \in \mathbb{C}^{M \times N}$ and $\mathbf{y} \in \mathbb{C}^{N}$ are uncorrelated, $\mathsf{vec}(\mathbf{X})$ and $\mathbf{y}$ have zero means and the correlation matrices $\mathbf{R}_{\mathrm{X}}$ and $\mathbf{R}_{\mathrm{y}}$, respectively. Define $\mathbf{D} \triangleq \mathsf{Diag}(\mathbf{y})$, then, the value of $\mathbf{R}$ can be derived as follows
\begin{align}
\mathbf{R} &= \mathbb{E} \{ \mathsf{vec}(\mathbf{X}\mathbf{D})\mathsf{vec}^{\mathrm{H}}(\mathbf{X}\mathbf{D}) \} \notag\\
&= \mathbb{E}_{\mathbf{X}, \mathbf{y}} \{ (\mathbf{D}^{\mathrm{T}} \otimes \mathbf{I}_{M})\mathsf{vec}(\mathbf{X})\mathsf{vec}^{\mathrm{H}}(\mathbf{X})(\mathbf{D}^{\mathrm{*}} \otimes \mathbf{I}_{M}) \} \notag\\
&= \mathbb{E}_{\mathbf{y}} \{ (\mathbf{D}^{\mathrm{T}} \otimes \mathbf{I}_{M})\mathbb{E}_{\mathbf{X}} \{ \mathsf{vec}(\mathbf{X})\mathsf{vec}^{\mathrm{H}}(\mathbf{X}) \} (\mathbf{D}^{\mathrm{*}} \otimes \mathbf{I}_{M}) \} \notag\\
&= \mathbb{E}_{\mathbf{y}} \{ \mathsf{Diag}(\mathbf{y} \otimes \bm{1}_{M})\mathbf{R}_{\mathrm{X}}\mathsf{Diag}(\mathbf{y}^{*} \otimes \bm{1}_{M}) \} \notag\\
&= \mathbf{R}_{\mathrm{X}} \odot \mathbb{E}_{\mathbf{y}} \{ (\mathbf{y} \otimes \bm{1}_{M})(\mathbf{y} \otimes \bm{1}_{M})^{\mathrm{H}} \} \notag\\
&= \mathbf{R}_{\mathrm{X}} \odot \mathbb{E}_{\mathbf{y}} \{ (\mathbf{y}\mathbf{y}^{\mathrm{H}}) \otimes (\bm{1}_{M}\bm{1}_{M}^{\mathrm{T}}) \} \notag\\
&= \mathbf{R}_{\mathrm{X}} \odot (\mathbb{E}_{\mathbf{y}} \{ \mathbf{y}\mathbf{y}^{\mathrm{H}} \} \otimes \mathbbm{1}_{M}) = \mathbf{R}_{\mathrm{X}} \odot (\mathbf{R}_{\mathrm{y}} \otimes \mathbbm{1}_{M}),
\end{align}
where $\bm{1}_{M}$ represents an $M$-dimensional all-one vector. Hence, invoking (78), the first two expectations in (76) are given by

\vspace{-3mm}
{\small
\begin{align}
&\mathbb{E} \{ \mathsf{vec}(\mathbf{G}\mathsf{Diag}(\mathbf{h}_{k}))\mathsf{vec}^{\mathrm{H}}(\mathbf{G}\mathsf{Diag}(\mathbf{h}_{k})) \} \!\!=\!\! \mathbf{R}_{\mathrm{G}} \!\!\odot\!\! (\mathbf{R}_{\mathrm{h},k} \!\otimes\! \mathbbm{1}_{M_{\mathrm{B}}}),\\
&\mathbb{E} \{ \mathsf{vec}(\hat{\mathbf{G}}\mathsf{Diag}(\hat{\mathbf{h}}_{k}))\mathsf{vec}^{\mathrm{H}}(\hat{\mathbf{G}}\mathsf{Diag}(\hat{\mathbf{h}}_{k})) \} \!\!=\!\! \mathbf{C}_{\hat{\mathrm{G}}} \!\!\odot\!\! (\mathbf{C}_{\hat{\mathrm{h}},k} \!\otimes\! \mathbbm{1}_{M_{\mathrm{B}}}).
\end{align}}

\vspace{-4mm}
Next, utilizing the following identities that are easily verified
\begin{align}
&\mathbb{E} \{ \mathbf{g}\hat{\mathbf{g}}^{\mathrm{H}} \} = \mathbf{R}_{\mathrm{G}}\mathsf{A}_{\mathrm{G}}^{\mathrm{H}}(\bm{\Psi})\mathbf{W}_{\mathrm{G}},\\
&\mathbb{E} \{ \mathbf{h}_{k}\hat{\mathbf{h}}_{k}^{\mathrm{H}} \} = \mathbf{R}_{\mathrm{h},k}\bm{\Psi}^{\mathrm{*}}\mathbf{W}_{\mathrm{h},k}, \; k \in \mathcal{K},
\end{align}
the third expectation in (76) can be derived as follows
\begin{align}
&\mathbb{E} \{ \mathsf{vec}(\mathbf{G}\mathsf{Diag}(\mathbf{h}_{k}))\mathsf{vec}^{\mathrm{H}}(\hat{\mathbf{G}}\mathsf{Diag}(\hat{\mathbf{h}}_{k})) \} \notag\\
= \;&\mathbb{E}_{\mathbf{G}, \mathbf{h}} \{ (\mathbf{D}_{\mathrm{h},k}^{\mathrm{T}} \otimes \mathbf{I}_{M_{\mathrm{B}}})\mathsf{vec}(\mathbf{G})\mathsf{vec}^{\mathrm{H}}(\hat{\mathbf{G}})(\hat{\mathbf{D}}_{\mathrm{h},k}^{*} \otimes \mathbf{I}_{M_{\mathrm{B}}}) \} \notag\\
= \;&\mathbb{E}_{\mathbf{h}} \{ (\mathbf{D}_{\mathrm{h},k}^{\mathrm{T}} \otimes \mathbf{I}_{M_{\mathrm{B}}})\mathbb{E}_{\mathbf{G}} \{ \mathsf{vec}(\mathbf{G})\mathsf{vec}^{\mathrm{H}}(\hat{\mathbf{G}}) \} (\hat{\mathbf{D}}_{\mathrm{h},k}^{*} \otimes \mathbf{I}_{M_{\mathrm{B}}}) \} \notag\\
= \;&\mathbb{E}_{\mathbf{h}} \{ \mathsf{Diag}(\mathbf{h}_{k} \otimes \bm{1}_{M_{\mathrm{B}}})\mathbf{R}_{\mathrm{G}}\mathsf{A}_{\mathrm{G}}^{\mathrm{H}}(\bm{\Psi})\mathbf{W}_{\mathrm{G}}\mathsf{Diag}(\hat{\mathbf{h}}_{k}^{*} \otimes \bm{1}_{M_{\mathrm{B}}}) \} \notag\\
= \;&(\mathbf{R}_{\mathrm{G}}\mathsf{A}_{\mathrm{G}}^{\mathrm{H}}(\bm{\Psi})\mathbf{W}_{\mathrm{G}}) \odot \mathbb{E}_{\mathbf{h}} \{ (\mathbf{h}_{k} \otimes \bm{1}_{M_{\mathrm{B}}})(\hat{\mathbf{h}}_{k} \otimes \bm{1}_{M_{\mathrm{B}}})^{\mathrm{H}} \} \notag\\
= \;&(\mathbf{R}_{\mathrm{G}}\mathsf{A}_{\mathrm{G}}^{\mathrm{H}}(\bm{\Psi})\mathbf{W}_{\mathrm{G}}) \odot \mathbb{E}_{\mathbf{h}} \{ (\mathbf{h}_{k}\hat{\mathbf{h}}_{k}^{\mathrm{H}}) \otimes (\bm{1}_{M_{\mathrm{B}}}\bm{1}_{M_{\mathrm{B}}}^{\mathrm{T}}) \} \notag\\
= \;&(\mathbf{R}_{\mathrm{G}}\mathsf{A}_{\mathrm{G}}^{\mathrm{H}}(\bm{\Psi})\mathbf{W}_{\mathrm{G}}) \odot (\mathbb{E}_{\mathbf{h}} \{ (\mathbf{h}_{k}\hat{\mathbf{h}}_{k}^{\mathrm{H}}) \} \otimes \mathbbm{1}_{M_{\mathrm{B}}}) \notag\\
= \;&(\mathbf{R}_{\mathrm{G}}\mathsf{A}_{\mathrm{G}}^{\mathrm{H}}(\bm{\Psi})\mathbf{W}_{\mathrm{G}}) \odot ((\mathbf{R}_{\mathrm{h},k}\bm{\Psi}^{\mathrm{*}}\mathbf{W}_{\mathrm{h},k}) \otimes \mathbbm{1}_{M_{\mathrm{B}}}),
\end{align}
where $\mathbf{D}_{\mathrm{h},k} \triangleq \mathsf{Diag}(\mathbf{h}_{k}), \; k \in \mathcal{K}$, and $\hat{\mathbf{D}}_{\mathrm{h},k} \triangleq \mathsf{Diag}(\hat{\mathbf{h}}_{k}), \; k \in \mathcal{K}$. Combining (79), (80), (83) and its hermitian transpose, we obtain (10).

\subsection*{B. Derivation of $\nabla_{\bm{\Theta}}\mathsf{g}(\bm{\Theta})$}

Define $\tilde{\mathsf{g}}(\bm{\Phi})$ as the objective of $(\mathcal{P}2)$ and we firstly calculate $\frac{\partial}{\partial\bm{\Phi}}\tilde{\mathsf{g}}(\bm{\Phi})$. It can be seen that the Hadamard product and the Kronecker product make it difficult to derive. Notice that $\mathsf{Tr} \{ \mathbf{A} \odot \mathbf{B} \}$ can be written as
\begin{equation}
\mathsf{Tr} \{ \mathbf{A} \odot \mathbf{B} \} = \mathsf{Tr} \{ \mathsf{Ddiag}(\mathbf{A})\mathbf{B} \} = \mathsf{Tr} \{ \mathbf{A}\mathsf{Ddiag}(\mathbf{B}) \}.
\end{equation}
Via exploiting (84), the Hadamard product in (12) disappears.

To overcome the difficulty led by the Kronecker product, we first define and study the following derivatives
\begin{align}
&\mathbf{F}_{\mathrm{o}}(\mathbf{A}, \mathbf{B}, \mathbf{C}, \mathbf{X}) \triangleq \frac{\partial}{\partial\mathbf{X}}\mathsf{Tr} \{ \mathbf{C}((\mathbf{AXB}) \otimes \mathbbm{1}_{N}) \}, \\
&\mathbf{F}_{\mathrm{e}}(\mathbf{A}, \mathbf{B}, \mathbf{C}, \mathbf{X}) \triangleq \frac{\partial}{\partial\mathbf{X}}\mathsf{Tr} \{ \mathbf{C}((\mathbf{AXB}) \otimes \mathbf{I}_{N}) \},
\end{align}
where the matrices $\mathbf{A}$, $\mathbf{B}$ and $\mathbf{X}$ are $M$-dimensional complex square matrices and $\mathbf{C} \in \mathbb{C}^{MN \times MN}$.

We firstly take (85) into consideration. Denote an $M$-dimensional matrix $\tilde{\mathbf{C}}$ whose $(i,j)$th element is given by $\sum_{s = (i - 1)N + 1}^{iN}\sum_{t = (j - 1)N + 1}^{jN}[\mathbf{C}]_{st}$. Then, $\mathsf{Tr} \{ \mathbf{C}((\mathbf{AXB}) \otimes \mathbbm{1}_{N}) \}$ in (85) can be calculated as
\begin{equation}
\mathsf{Tr} \{ \mathbf{C}((\mathbf{AXB}) \otimes \mathbbm{1}_{N}) \} = \sum_{m=1}^{M}[\tilde{\mathbf{C}}]_{m,:}\mathbf{AX}[\mathbf{B}]_{:,m}.
\end{equation}
Hence, the derivative shown in (85) reads as
\begin{align}
&\frac{\partial}{\partial\mathbf{X}}\mathsf{Tr} \{ \mathbf{C}((\mathbf{AXB}) \otimes \mathbbm{1}_{N}) \} = \frac{\partial}{\partial\mathbf{X}}\sum_{m=1}^{M}[\tilde{\mathbf{C}}]_{m,:}\mathbf{AX}[\mathbf{B}]_{:,m} \notag\\
= &\frac{\partial}{\partial\mathbf{X}}\mathsf{Tr} \bigg\{ \sum_{m=1}^{M}[\mathbf{B}]_{:,m}[\tilde{\mathbf{C}}]_{m,:}\mathbf{AX} \bigg\} = (\mathbf{B}\tilde{\mathbf{C}}\mathbf{A})^{\mathrm{T}}.
\end{align}

Following the similar procedure, function (86) has the same form as (85), where the $(i,j)$th element of $\tilde{\mathbf{C}}$ is $\mathsf{Tr} \{ [\mathbf{C}]_{(i - 1)N + 1:iN,(j - 1)N + 1:jN} \}$ instead.

Via leveraging (85) and (86), the derivative $\frac{\partial}{\partial\bm{\Phi}}\tilde{\mathsf{g}}(\bm{\Phi})$ can be expressed as
\begin{equation}
\frac{\partial}{\partial\bm{\Phi}}\tilde{\mathsf{g}}(\bm{\Phi}) = \sum_{i=1}^{2}\mathbf{D}_{\bm{\Phi},i}^{\mathrm{o}} + \mathbf{D}_{\bm{\Phi}}^{\mathrm{e}},
\end{equation}
where

\vspace{-4mm}
{\small
\begin{align}
&\mathbf{D}_{\bm{\Phi},1}^{\mathrm{o}} \triangleq {\sum}_{k=1}^{K}\mathbf{F}_{\mathrm{o}}(\mathbf{W}_{\mathrm{h},k}^{\mathrm{H}}\mathbf{P}, \mathbf{R}_{\mathrm{h},k}(\bm{\Phi}\mathbf{P})^{*}\mathbf{W}_{\mathrm{h},k}, \mathsf{Ddiag}(\mathbf{C}_{1}^{\mathrm{o}}), \bm{\Phi}), \notag\\
&\mathbf{D}_{\bm{\Phi},2}^{\mathrm{o}} \triangleq -{\sum}_{k=1}^{K}\mathbf{F}_{\mathrm{o}}(\mathbf{W}_{\mathrm{h},k}^{\mathrm{H}}\mathbf{P}, \mathbf{R}_{\mathrm{h},k}, \mathsf{Ddiag}(\mathbf{C}_{2}^{\mathrm{o}}), \bm{\Phi}), \notag\\
&\mathbf{D}_{\bm{\Phi}}^{\mathrm{e}} \triangleq \mathbf{F}_{\mathrm{e}}(\mathbf{P}, \mathbf{I}_{M_{\mathrm{R}}}, \mathbf{C}_{1}^{\mathrm{e}} + \mathbf{C}_{2}^{\mathrm{e}} - \mathbf{C}_{3}^{\mathrm{e}}, \bm{\Phi}), \notag\\
&\mathbf{C}_{1}^{\mathrm{o}} \triangleq \mathbf{W}_{\mathrm{G}}^{\mathrm{H}}\mathbf{C}_{1,1}^{\mathrm{o}}\mathbf{W}_{\mathrm{G}}, \notag\\
&\mathbf{C}_{1,1}^{\mathrm{o}} \triangleq ((\bm{\Phi}\mathbf{P})^{\mathrm{T}} \otimes \mathbf{I}_{M_{\mathrm{B}}})\mathbf{R}_{\mathrm{G}}((\bm{\Phi}\mathbf{P})^{\mathrm{*}} \otimes \mathbf{I}_{M_{\mathrm{B}}}) + \bm{\Sigma}_{\mathrm{B}} \otimes \mathbf{I}_{M_{\mathrm{R}}}, \notag\\
&\mathbf{C}_{2}^{\mathrm{o}} \triangleq \mathbf{W}_{\mathrm{G}}^{\mathrm{H}}((\bm{\Phi}\mathbf{P})^{\mathrm{T}} \otimes \mathbf{I}_{M_{\mathrm{B}}})\mathbf{R}_{\mathrm{G}}, \notag\\
&\mathbf{C}_{1}^{\mathrm{e}} \triangleq \mathbf{R}_{\mathrm{G}}((\bm{\Phi}\mathbf{P})^{\mathrm{*}} \otimes \mathbf{I}_{M_{\mathrm{B}}})\mathbf{W}_{\mathrm{G}}\mathbf{C}_{1,1}^{\mathrm{e}}\mathbf{W}_{\mathrm{G}}^{\mathrm{H}}, \notag\\
&\mathbf{C}_{2}^{\mathrm{e}} \triangleq \mathbf{R}_{\mathrm{G}}((\bm{\Phi}\mathbf{P})^{\mathrm{*}} \otimes \mathbf{I}_{M_{\mathrm{B}}})\mathbf{W}_{\mathrm{G}}\mathbf{C}_{2,1}^{\mathrm{e}}\mathbf{W}_{\mathrm{G}}^{\mathrm{H}}, \notag\\
&\mathbf{C}_{3}^{\mathrm{e}} \triangleq \mathbf{R}_{\mathrm{G}}\mathbf{C}_{3,1}^{\mathrm{e}}\mathbf{W}_{\mathrm{G}}^{\mathrm{H}}, \notag\\
&\mathbf{C}_{1,1}^{\mathrm{e}} \triangleq {\sum}_{k=1}^{K}\mathsf{Ddiag}((\mathbf{W}_{\mathrm{h},k}^{\mathrm{H}}(\bm{\Phi}\mathbf{P})^{\mathrm{T}}\mathbf{R}_{\mathrm{h},k}(\bm{\Phi}\mathbf{P})^{*}\mathbf{W}_{\mathrm{h},k}) \otimes \mathbbm{1}_{M_{\mathrm{B}}}), \notag\\
&\mathbf{C}_{2,1}^{\mathrm{e}} \triangleq {\sum}_{k=1}^{K}\mathsf{Ddiag}((\mathbf{W}_{\mathrm{h},k}^{\mathrm{H}}\bm{\Sigma}_{\mathrm{U},k}\mathbf{W}_{\mathrm{h},k}) \otimes \mathbbm{1}_{M_{\mathrm{B}}}), \notag\\
&\mathbf{C}_{3,1}^{\mathrm{e}} \triangleq {\sum}_{k=1}^{K}\mathsf{Ddiag}((\mathbf{W}_{\mathrm{h},k}^{\mathrm{H}}(\bm{\Phi}\mathbf{P})^{\mathrm{T}}\mathbf{R}_{\mathrm{h},k}) \otimes \mathbbm{1}_{M_{\mathrm{B}}}).
\end{align}}

\vspace{-4mm}
Moreover, recalling $\bm{\Phi} = e^{j\bm{\Theta}}$ and utilizing the chain rule, we have
\begin{equation}
\frac{\partial}{\partial\bm{\Theta}}\mathsf{g}(\bm{\Theta}) = j\frac{\partial}{\partial\bm{\Phi}}\tilde{\mathsf{g}}(\bm{\Phi}) \odot e^{j\bm{\Theta}}.
\end{equation}
Due to $\mathsf{g}(\bm{\Theta})$ real-valued, $\nabla_{\bm{\Theta}}\mathsf{g}(\bm{\Theta})$ can be calculated as [39]
\begin{align}
&\nabla_{\bm{\Theta}}\mathsf{g}(\bm{\Theta}) = \frac{\partial}{\partial\bm{\Theta}}\mathsf{g}(\bm{\Theta}) + \frac{\partial}{\partial\bm{\Theta}^{*}}\mathsf{g}(\bm{\Theta}) \notag\\
&= \frac{\partial}{\partial\bm{\Theta}}\mathsf{g}(\bm{\Theta}) + \bigg(\frac{\partial}{\partial\bm{\Theta}}\mathsf{g}(\bm{\Theta})\bigg)^{*} = 2\mathsf{Re} \bigg\{ \frac{\partial}{\partial\bm{\Theta}}\mathsf{g}(\bm{\Theta}) \bigg\}.
\end{align}

\subsection*{C. Derivation of $\nabla_{\mathbf{p}}\mathsf{h}(\mathbf{p})$}

Denote $\tilde{\mathsf{h}}(\mathbf{P})$ (recall that $\mathbf{P} = \mathsf{Diag}(\mathbf{p})$) as the objective of $(\mathcal{P}4)$ and we firstly investigate $\nabla_{\mathbf{P}}\tilde{\mathsf{h}}(\mathbf{P})$. Similar to the procedure shown in Appendix B, $\nabla_{\mathbf{P}}\tilde{\mathsf{h}}(\mathbf{P})$ can be given by
\begin{equation}
\nabla_{\mathbf{P}}\tilde{\mathsf{h}}(\mathbf{P}) = 2\mathsf{Re} \bigg\{ \sum_{i=1}^{2}\mathbf{D}_{\mathbf{P},i}^{\mathrm{o}} + \mathbf{D}_{\mathbf{P}}^{\mathrm{e}} \bigg\},
\end{equation}
where

\vspace{-4mm}
{\small
\begin{align}
&\mathbf{D}_{\mathbf{P},1}^{\mathrm{o}} \triangleq {\sum}_{k=1}^{K}\mathbf{F}_{\mathrm{o}}(\mathbf{W}_{\mathrm{h},k}^{\mathrm{H}}, \bm{\Phi}^{\mathrm{T}}\mathbf{R}_{\mathrm{h},k}(\bm{\Phi}\mathbf{P})^{*}\mathbf{W}_{\mathrm{h},k}, \mathsf{Ddiag}(\mathbf{C}_{1}^{\mathrm{o}}), \mathbf{P}), \notag\\
&\mathbf{D}_{\mathbf{P},2}^{\mathrm{o}} \triangleq -{\sum}_{k=1}^{K}\mathbf{F}_{\mathrm{o}}(\mathbf{W}_{\mathrm{h},k}^{\mathrm{H}}, \bm{\Phi}^{\mathrm{T}}\mathbf{R}_{\mathrm{h},k}, \mathsf{Ddiag}(\mathbf{C}_{2}^{\mathrm{o}}), \mathbf{P}), \notag\\
&\mathbf{D}_{\mathbf{P}}^{\mathrm{e}} \triangleq \mathbf{F}_{\mathrm{e}}(\mathbf{I}_{M_{\mathrm{R}}}, \bm{\Phi}^{\mathrm{T}}, \mathbf{C}_{1}^{\mathrm{e}} + \mathbf{C}_{2}^{\mathrm{e}} - \mathbf{C}_{3}^{\mathrm{e}}, \mathbf{P}),
\end{align}}

\vspace{-5mm}
\noindent and the notations in (94) have been defined in (90).

By exploiting $\mathbf{p} = \mathsf{diag}(\mathbf{P})$, we obtain
\begin{equation}
\nabla_{\mathbf{p}}\mathsf{h}(\mathbf{p}) = \mathsf{diag}(\nabla_{\mathbf{P}}\tilde{\mathsf{h}}(\mathbf{P})).
\end{equation}

\subsection*{D. Proof of Theorem 2}

\begin{proof}

Consider the following optimization
\begin{align}
(\mathcal{P}11): \min_{\mathbf{p} \in \mathbb{R}^{N}} \; &\| \mathbf{p} \|_{2}^{2} - 2\mathsf{Re} \{ \mathbf{q}^{\mathrm{H}}\mathbf{p} \} \\
\mathrm{s.t.} \; &\| \mathbf{p} \|_{2}^{2} \leq P, \tag{96a}\\
&[\mathbf{p}]_{n} \geq 0, \; n = 1, \cdots, N, \tag{96b}
\end{align}
which is strictly convex.

The Lagrangian function of $(\mathcal{P}11)$ reads as
\begin{equation}
\mathcal{L}(\mathbf{p},\bm{\lambda},\mu) = \| \mathbf{p} \|_{2}^{2} - 2\mathsf{Re} \{ \mathbf{q}^{\mathrm{H}}\mathbf{p} \} - \bm{\lambda}^{\mathrm{T}}\mathbf{p} + \mu(\| \mathbf{p} \|_{2}^{2} - P).
\end{equation}
By setting $\frac{\partial\mathcal{L}(\mathbf{p},\bm{\lambda},\mu)}{\partial\mathbf{p}}$ to $0$, we obtain
\begin{equation}
\mathbf{p} = \frac{1}{1 + \mu}\bigg(\mathsf{Re} \{ \mathbf{q} \} + \frac{1}{2}\bm{\lambda}\bigg).
\end{equation}

According to the KKT condition
\begin{align}
&[\bm{\lambda}]_{n} \geq 0, \; [\bm{\lambda}]_{n}[\mathbf{p}]_{n} = 0, \; n = 1, \cdots, N,\\
&\mu \geq 0, \; \mu(\| \mathbf{p} \|_{2}^{2} - P) = 0,
\end{align}
and constraint (96b), we can conclude from (98) that
\begin{itemize}

\item if $[\mathsf{Re} \{ \mathbf{q} \}]_{n} \leq 0$, we have $[\bm{\lambda}]_{n} \geq 0$, $[\mathbf{p}]_{n} = 0$,

\item otherwise, $[\mathbf{p}]_{n} > 0$, $[\bm{\lambda}]_{n} = 0$.

\end{itemize}
Therefore, (98) can be simplified as
\begin{equation}
\mathbf{p} = \frac{[\mathsf{Re} \{ \mathbf{q} \}]_{+}}{1 + \mu},
\end{equation}
where $[\mathbf{x}]_{+} \triangleq \max \{ \mathbf{x}, \bm{0} \}$ in an element-wise manner.

Following (100), we can observe that
\begin{itemize}

\item if $\| [\mathsf{Re} \{ \mathbf{q} \}]_{+} \|_{2}^{2} \leq P$ when $\mu = 0$, we obtain
\begin{equation}
\mathbf{p}^{\star} = [\mathsf{Re} \{ \mathbf{q} \}]_{+},
\end{equation}

\item otherwise, there must exist $\mu > 0$ such that
\begin{equation}
\mathbf{p}^{\star} = \frac{\sqrt{P}}{\| [\mathsf{Re} \{ \mathbf{q} \}]_{+} \|_{2}}[\mathsf{Re} \{ \mathbf{q} \}]_{+},
\end{equation}

\end{itemize}
Substituting the parameters in $(\mathcal{P}5)$ into (102) or (103), we obtain the desired result.
\end{proof}

\subsection*{E. Derivation of (20)}

The derivative $\frac{\partial}{\partial\mathbf{W}_{\mathrm{G}}^{*}}\sum_{k=1}^{K}\mathsf{Tr} \{ \mathbf{C}_{\mathrm{H}_{\mathrm{c},k}}(\bm{\Phi}, \mathbf{P}, \mathbf{W}_{\mathrm{G}}, \bm{\mathcal{W}}_{\mathrm{h}}) \}$ can be written as
\begin{align}
&\frac{\partial}{\partial\mathbf{W}_{\mathrm{G}}^{*}}\sum_{k=1}^{K}\mathsf{Tr} \{ \mathbf{C}_{\mathrm{H}_{\mathrm{c},k}}(\bm{\Phi}, \mathbf{P}, \mathbf{W}_{\mathrm{G}}, \bm{\mathcal{W}}_{\mathrm{h}}) \} = (\mathsf{A}_{\mathrm{G}}(\bm{\Psi})\mathbf{R}_{\mathrm{G}}\mathsf{A}_{\mathrm{G}}^{\mathrm{H}}(\bm{\Psi}) \notag\\
&+ \bm{\Sigma}_{\mathrm{B}} \otimes \mathbf{I}_{M_{\mathrm{R}}})\mathbf{W}_{\mathrm{G}}{\sum}_{k=1}^{K}\mathsf{Ddiag}(\mathbf{C}_{\hat{\mathrm{h}},k} \otimes \mathbbm{1}_{M_{\mathrm{B}}}) - \mathsf{A}_{\mathrm{G}}(\bm{\Psi})\mathbf{R}_{\mathrm{G}} \notag\\
&\times {\sum}_{k=1}^{K}\mathsf{Ddiag}((\mathbf{W}_{\mathrm{h},k}^{\mathrm{H}}\bm{\Psi}^{\mathrm{T}}\mathbf{R}_{\mathrm{h},k}) \otimes \mathbbm{1}_{M_{\mathrm{B}}}),
\end{align}
where (84) is utilized. Then, by setting the above derivative to zero and exploiting (21), we obtain (20).

\end{document}